\documentclass[aps,prd,reprint,superscriptaddress,nofootinbib,floatfix,preprintnumbers]{revtex4-2}

\usepackage{xcolor}
\usepackage{graphicx}
\usepackage{amsmath}
\usepackage{amsfonts}
\usepackage{amssymb}
\usepackage{caption}

\usepackage[colorlinks=true, linkcolor=blue, citecolor=blue, urlcolor=blue]{hyperref}

\begin{document}

\title{Enhanced Signal and Excited-State Control in Nonlocal Lattice Correlators}

\author{Shi-Yi Zhong}
\affiliation{School of Science and Engineering, The Chinese University of Hong Kong, Shenzhen, Guangdong, 518172, P.R. China}

\author{Jian-Hui Zhang}
\affiliation{School of Science and Engineering, The Chinese University of Hong Kong, Shenzhen, Guangdong, 518172, P.R. China}
\affiliation{Southern Center for Nuclear-Science Theory (SCNT), Institute of Modern Physics, Chinese Academy of Sciences, Huizhou, Guangdong 516000, China}

\author{Dian-Jun Zhao}
\affiliation{School of Science and Engineering, The Chinese University of Hong Kong, Shenzhen, Guangdong, 518172, P.R. China}

\begin{abstract}
Nonlocal correlation functions underpin lattice calculations of parton physics. However, their extraction is limited by a rapidly deteriorating signal-to-noise ratio with increasing Euclidean separation and hadron momentum, and by excited-state contamination. We propose a new fitting strategy that enhances control over both: by taking the ratio of nonlocal to local correlators, we can project out the dominant ground-state dynamics with a signal that is much cleaner and far less sensitive to excited states than the conventional approach. We demonstrate the method on the quark and gluon PDF, the quark GPD and TMDPDF. The results are consistent with those from the conventional three-point/two-point ratio fit but exhibit markedly improved precision and stability, and extend the accessible range to larger hadron momenta. 
For nonforward matrix elements such as GPDs, our strategy greatly simplifies the analysis procedure. 
Our method applies to general Dirac and momentum structures, for connected and disconnected contributions, and extends naturally to other nonlocal correlators, making it immediately deployable in existing lattice programs toward precision hadron structure. 
\end{abstract}

\maketitle

\section{Introduction}
\label{sec:intro}

Understanding how the quark and gluon degrees of freedom organize themselves inside a hadron is a central goal of modern nuclear and particle physics. This structure is encoded in parton distribution functions (PDFs), generalized parton distributions (GPDs), transverse-momentum-dependent distributions (TMDs), etc. These distributions provide essential nonperturbative inputs for the physics programs at Jefferson Lab, RHIC, and the Electron-Ion Colliders in the US and China~\cite{Accardi:2012qut,AbdulKhalek:2021gbh,Anderle:2021wcy} and are conventionally determined by a global fit to experimental data. On the other hand, lattice QCD offers a complementary approach to determining them from first principles, though the extraction is far from straightforward. The field-theoretic definitions of parton distributions are given on the lightcone, whereas lattice QCD is formulated in Euclidean spacetime and can only evaluate equal-time correlators. The traditional local operator approach accesses only the lowest few Mellin moments in practice, since moments of higher rank suffer from increasingly severe power-divergent operator mixing with lower-dimensional operators~\cite{Martinelli:1994ty,Detmold:2005gg,Liu:1993cv}.

Large-momentum effective theory (LaMET)~\cite{Ji:2013dva,Ji:2014gla,Ji:2020ect,Ji:2024oka} circumvents this obstruction by computing instead a \emph{quasi}-light-front (quasi-LF) correlation: a time-independent spatial correlation of a boosted hadron, defined from a nonlocal quark or gluon bilinear joined by a Wilson line. Taking the unpolarized quark PDF as an example, one computes the matrix element of the operator $\mathcal{O}^{\Gamma}(z)=\bar{q}(x)\Gamma W(x,x+z\hat{z})q(x+z\hat{z})$, with the Dirac structure $\Gamma$ set to $\gamma_t$. 
After Fourier transform, the momentum space quasi-distribution is related to its lightcone counterpart by a perturbative matching kernel and a systematic expansion in the hadron momentum. In this way, parton physics becomes accessible pointwise at moderate-$x$. Similar constructions also apply to the GPDs and TMDs. 
Related but distinct frameworks, such as the short-distance factorization approach of Refs.~\cite{Radyushkin:2017cyf,Radyushkin:2017lvu} and the current-current correlator construction of Refs.~\cite{Braun:2007wv,Ma:2017pxb,Chambers:2017dov}, share the same essential feature: the lightcone physics is encoded in the dependence of a nonlocal hadronic matrix element on the spatial separation $z$.

The computation of such a nonlocal matrix element is numerically demanding. In practice, it is extracted from a three-point correlation function $C_3(t,\tau,z)$, in which an interpolating field for the hadron is created at time zero, the nonlocal operator is inserted at time $\tau$, and the hadron is annihilated at the source--sink separation $t$. The matrix element is obtained from the large-$t$, fixed-$\tau$ behaviour of $C_3$, but the convergence to the ground-state plateau is controlled by the energy gaps to the excited states of the same quantum numbers, $E_n-E_0$, which are of the order a few hundred MeV for the pion and comparable to or smaller than the momentum of a boosted nucleon. The resulting excited-state contamination is, together with the renormalization of the nonlocal operator~\cite{Ji:2017oey,Ishikawa:2017iym,Green:2017xeu}, one of the two dominant systematic uncertainties of present-day lattice parton physics.

A variety of strategies has been developed to control the so-called excited-state contamination, and they differ substantially in cost, complexity, and the assumptions they require. The most widely used is the ratio method, in which $C_3$ is divided by the two-point function $C_2$ of the source interpolating field and the resulting ratio is fitted to a truncated spectral decomposition, with the number of retained states treated as a model choice. The unwanted excited-state contributions are controlled only by the explicitly modeled terms, and the truncation error is estimated either from a scan over the number of states and the fit windows or from a Bayesian model average over the candidate models~\cite{LatticeParton:2018gjr,Alexandrou:2019lfo}.
The \emph{summation} (or summed-ratio) method averages the three-point-to-two-point correlator ratio, $R(t,\tau)$, over $\tau$ before extracting the matrix element~\cite{Ma:2017pxb,Ma:2018bpg}, which removes part of the excited-state contribution at the price of a modified power correction, but it still relies on $C_2$ and therefore still requires a separate treatment of the two-point spectral content, and the $\tau$ summation converts the exponential excited-state contamination into a polynomial form~\cite{Bouchard:2016heu}.
The \emph{Feynman--Hellmann} (FH) theorem~\cite{Bouchard:2016heu} is closely related to the summed-ratio method. In its lattice implementation, the estimator is the same summed ratio, so the matrix element is obtained from a finite difference, rather than the value, of $R(t)$. It still carries the full $C_2$ normalization and hence the same overlap-ratio systematic as the conventional ratio.
\emph{Variational} approaches proceed differently: the generalized eigenvalue problem (GEVP) diagonalizes a matrix of two-point functions to construct optimized interpolating operators for the low-lying states~\cite{Luscher:1990ux,Blossier:2009kd}, after which the matrix element can be read off from the GEVP-projected three-point function with reduced overlap onto the excited states. It requires a basis of interpolating operators, a matrix of two-point functions, and (in the standard implementation) additional inversions and a careful treatment of the eigenvector contamination, and the statistical cost grows rapidly with the size of the basis.
\emph{Lanczos}-based methods reconstruct the spectral decomposition of the transfer matrix from Krylov-space moments of the correlators and thereby expose individual eigenstates and their matrix elements without a variational basis~\cite{Wagman:2024rid,Hackett:2024xnx,Hackett:2024nbe}; they provide a controlled spectral picture, but the analysis is still being developed.
Finally, direct \emph{multi-state fits with Bayesian priors} and model averaging are common in current analyses; their systematic uncertainty is the spread of the posterior over the truncation choices, which is difficult to shrink once the source--sink separations available on a given ensemble have been exhausted~\cite{Chen:2025cxr}. When the candidate models are enumerated explicitly, the truncation uncertainty can be quantified and propagated by Bayesian model averaging~\cite{Jay:2020ghd,Neil:2022jmn}, which converts model choice into a statistical statement at the cost of a larger set of fits.

In this paper, we propose a new strategy for fitting nonlocal operator matrix elements that exploits the structural features of nonlocal operators to significantly improve the signal-to-noise ratio and suppress excited-state contamination. We illustrate the method with quasi-LF correlators, but it applies equally to other nonlocal correlators. The central idea is fairly simple: instead of forming the conventional ratio of three-point correlators to two-point correlators, we form the ratio of nonlocal to local three-point correlators, $C_3(t,\tau,z)/C_3(t,\tau,0)$. This construction enables a cleaner extraction of ground-state information, with substantially reduced sensitivity to excited-state contributions. We call this the \emph{self-ratio} and derive its spectral decomposition in Sec.~\ref{sec:theory}. Because the numerator and denominator share the same spectrum, overlaps, and source--sink structure, the overlap factors and the energy denominators cancel identically between them, and the self-ratio reduces to a ratio of two spectral sums of operator matrix elements. The residual contamination is therefore governed by the operator rather than by the interpolating field, in contrast to the conventional ratio, whose residual is governed by the overlap ratio $\zeta_1/\zeta_0$, where $\zeta_n\equiv\langle n|\chi|\Omega\rangle$ is the overlap with the $n$-th state, and $|\Omega\rangle$ is the vacuum.
The cancellation requires only that the numerator and the denominator describe the same transition, with identical interpolating fields, momenta, and time arguments. It holds for arbitrary kinematics, degenerate or not, and for real overlaps, so the self-ratio method also applies to non-forward kinematics required by the GPD calculation, where the source and sink momenta differ. The $z=0$ correlator is already part of the standard correlator set, so the method adds no extra measurement. 
We test the method on four channels: the isovector quark transversity and the unpolarized gluon PDFs of the nucleon, the pion GPD, and the nucleon Boer--Mulders function.

The remainder of this paper is organized as follows. Section~\ref{sec:theory} develops the theoretical framework: taking the quasi-LF correlators as an example, we derive the spectral decomposition of the self-ratio, examine the spectral structure of its residual contamination, describe how that residual can be measured and modeled, and extend the construction to non-forward kinematics and to transverse-momentum-dependent distributions. We then compare the method with existing approaches and state its limitations. Section~\ref{sec:case} presents the case studies. Section~\ref{sec:summary} contains a summary and outlook.

\section{Theoretical Framework}
\label{sec:theory}
In this section, we discuss the theoretical foundation of the method, taking the quasi-LF correlator as an example. However, the discussion can be easily extended to other nonlocal operator matrix elements, such as current-current correlators and those probing multiparton physics~\cite{Zhang:2023wea,Jaarsma:2023woo,Reitinger:2024ulw,Diehl:2025kcr}.

\subsection{Definitions and conventions}
\label{sec:defs}

We work in Euclidean spacetime on a lattice of temporal extent $L_t$ and spatial extent $L_s^3$ with lattice spacing $a$. Let $\chi(\vec{x},t)$ be an interpolating field for the hadron, and let
\begin{equation}
  \mathcal{O}^{\Gamma}(z;\vec{y},\tau)
  = \bar{q}(\vec{y},\tau)\,\Gamma\,
  W\left[(\vec{y},\tau),(\vec{y}+z\hat{z},\tau)\right]q(\vec{y}+z\hat{z},\tau)
  \label{eq:op}
\end{equation}
be the nonlocal quark bilinear with a straight Wilson line of length $z$ along the spatial $\hat{z}$ direction. The Dirac structure $\Gamma$ selects the observable: $\gamma_t$, $\gamma_t\gamma_5$, or $\sigma_{t\perp}\gamma_5$ ($\perp=x,y$) for the unpolarized, helicity, and transversity cases.

Let $|n\rangle$ be the eigenstates of the lattice transfer matrix with definite momentum, $H|n\rangle=E_n|n\rangle$. 
We take the overlaps real; for complex overlaps the corresponding conjugations are understood and the cancellation below is unaffected. The two-point function is
\begin{equation}
  C_2(t) = \sum_{n} \frac{\zeta_n^2}{2E_n}
  \left(e^{-E_n t}+e^{-E_n(L_t-t)}\right) ,
  \label{eq:c2}
\end{equation}
and, for $0<\tau<t$ and neglecting contact terms in which the operator contracts with the source or sink,
\begin{equation}
  C_3(t,\tau,z) = \sum_{m,n}
  \frac{O_{mn}(z)\,\zeta_m \zeta_n}{(2E_m)(2E_n)}
  \,e^{-E_m(t-\tau)}\,e^{-E_n\tau} ,
  \label{eq:c3}
\end{equation}
where $O_{mn}(z)\equiv\langle m|\mathcal{O}^{\Gamma}(z)|n\rangle$; the ground-state matrix element is $h_B(z)\equiv O_{00}(z)$, whose Fourier transform gives the quasi-distribution.

We compare two normalizations, the conventional ratio
\begin{equation}
  R(t,\tau,z) \equiv \frac{C_3(t,\tau,z)}{C_2(t)}
  \label{eq:ratio}
\end{equation}
and the self-ratio
\begin{equation}
  R_{\rm self}(t,\tau,z) \equiv \frac{C_3(t,\tau,z)}{C_3(t,\tau,0)} ,
  \label{eq:selfratio}
\end{equation}
taken at the same insertion time, source--sink separation and hadron momenta, with $D_{mn}\equiv\langle m|\mathcal{O}^{\Gamma}(0)|n\rangle$. The fitted ground-state parameter of the self-ratio is
\begin{equation}
  \hat h_B(z) \equiv \frac{h_B(z)}{h_B(0)} = \frac{O_{00}(z)}{D_{00}} .
  \label{eq:hatdef}
\end{equation}
In general $D_{00}=h_B(0)$ is the local charge (or the elastic form factor in the non-forward case) and must be supplied separately; see Sec.~\ref{sec:assessment}.

\subsection{Spectral structure of self-ratio and the first-order approximation}
\label{sec:selfratio}

We begin with initial and final states of equal momentum magnitude, so that the two spectra coincide; this covers the forward case $\vec{P}_f=\vec{P}_i$ and the Breit frame $\vec{P}_f=-\vec{P}_i$ in the non-forward case. Dividing Eq.~\eqref{eq:c3} by its $z=0$ counterpart, the overlap factors, energy denominators and source--sink exponentials are identical in numerator and denominator, because both describe the same transition with the same interpolating fields, momenta and time arguments. They cancel term by term, and
\begin{equation}
  R_{\rm self}(t,\tau,z)
  = \frac{\displaystyle\sum_{m,n}
  \frac{O_{mn}(z)}{(2E_m)(2E_n)}\,\zeta_m\zeta_n\,
  e^{-E_m(t-\tau)}e^{-E_n\tau}}
  {\displaystyle\sum_{m,n}
  \frac{D_{mn}}{(2E_m)(2E_n)}\,\zeta_m\zeta_n\,
  e^{-E_m(t-\tau)}e^{-E_n\tau}} ,
  \label{eq:selfratio-exact}
\end{equation}
with no dependence on the interpolating field beyond the relative weights of the excited-state terms. With $\Delta E_n\equiv E_n-E_0$ and
\begin{align}
  &\Sigma(z) \equiv \sum_{(m,n)\neq(0,0)}
  \frac{O_{mn}(z)}{O_{00}(z)}\,\omega_{mn} ,\nonumber\\
  &\omega_{mn} \equiv \frac{\zeta_m\zeta_n}{\zeta_0^2}
  \frac{E_0^2}{E_mE_n}\,e^{-\Delta E_m(t-\tau)}e^{-\Delta E_n\tau},
  \label{eq:Sigma}
\end{align}
the ratio becomes
\begin{equation}
  R_{\rm self}(t,\tau,z)
  = \frac{h_B(z)}{h_B(0)}
  \frac{1+\Sigma(z)}{1+\Sigma(0)} .
  \label{eq:selfratio-ratio}
\end{equation}
To first order in the admixtures,
\begin{equation}
\begin{split}
  R_{\mathrm{self}}(t,\tau,z) ={}& \frac{h_B(z)}{h_B(0)}\Big[1
  + \sum_{(m,n)\neq(0,0)}
  \Big(\frac{O_{mn}(z)}{O_{00}(z)}\\
  &- \frac{D_{mn}}{D_{00}}\Big)\omega_{mn}
  + O(\omega^2)\Big] ,
\end{split}
  \label{eq:selfratio-firstorder}
\end{equation}
with
\begin{equation}
  c_{mn}(z) \equiv \frac{O_{mn}(z)}{O_{00}(z)} - \frac{D_{mn}}{D_{00}} .
  \label{eq:cmn}
\end{equation}
The coefficients vanish identically at $z=0$, because $O_{mn}(0)=D_{mn}$. 
In contrast with the commonly used $\tau$ cuts which shrink every excited-state term, the self-ratio cancels the part shared by the numerator and denominator and leaves a residual that vanishes at $z=0$ and turns on away from it. The conventional ratio makes this contrast explicit.

\subsection{Relation to the conventional ratio}
\label{sec:conventional}

To the same order, Eq.~\eqref{eq:ratio} gives
\begin{equation}
\begin{split}
  R(t,\tau,z) ={}& \frac{h_B(z)}{2E_0}f(t)\Big[
  1 + \sum_{(m,n)\neq(0,0)}\frac{O_{mn}(z)}{O_{00}(z)}\omega_{mn}\\
  &- \Pi(t) + O(\omega^2)\Big] ,
\end{split}
  \label{eq:ratio-firstorder}
\end{equation}
where $f(t)\equiv[1+e^{-E_0(L_t-2t)}]^{-1}$ accompanies the ratio because $C_2$ contains both temporal images while $C_3$ propagates only forward in $t$; it is $z$-independent and unity for $t\ll L_t/2$. The two-point contamination is
\begin{equation}
  \Pi(t) \equiv \sum_{n\neq0}
  \frac{\zeta_n^2}{\zeta_0^2}\frac{E_0}{E_n}
  \frac{e^{-E_nt}+e^{-E_n(L_t-t)}}{e^{-E_0t}+e^{-E_0(L_t-t)}} .
  \label{eq:Pi}
\end{equation}
Three differences follow. $-\Pi(t)$ appears only in the conventional ratio. Its operator contamination enters with the full $O_{mn}(z)/O_{00}(z)$, not the difference at $z$ and $0$. The coefficient of $\Pi(t)$ involves $(\zeta_n/\zeta_0)^2$, a property of the interpolating field that the three-point data cannot constrain. Writing $R_{\rm self}=\hat h_B[1+\varepsilon_{\rm self}]$ and $R=(h_B/2E_0)f(t)[1+\varepsilon_{\rm conv}]$, the two contaminations satisfy
\begin{equation}
  1+\varepsilon_{\rm self}(t,\tau,z)
  = \frac{1+\varepsilon_{\rm conv}(t,\tau,z)}
  {1+\varepsilon_{\rm conv}(t,\tau,0)} ,
  \label{eq:common-mode}
\end{equation}
or, to first order, $\varepsilon_{\rm self}=\varepsilon_{\rm conv}(z)-\varepsilon_{\rm conv}(0)$. The self-ratio is the difference of two conventional contaminations: it removes what the two separations share and nothing else. It is obvious that the gain is largest when the operator's excited-state structure varies slowly with $z$. 

\subsection{Spectral structure of the residual}
\label{sec:spectral}

The residual is governed by $c_{mn}(z)$, which carries neither $\Pi(t)$ nor the overlap ratio; the overlap ratio survives only in the weights $\omega_{mn}$ of Eq.~\eqref{eq:Sigma}. While the off-diagonal coefficients carry both the operator's $z$-dependence and the $\tau$-arch whose shape is fixed by the spectrum, the diagonal coefficients carry the operator's $z$-dependence only.

When $m=n$, Eq.~\eqref{eq:Sigma} reduces to
\begin{equation}
  \omega_{nn} = \frac{\zeta_n^2}{\zeta_0^2}\frac{E_0^2}{E_n^2}
  e^{-\Delta E_n t} ,
  \label{eq:wnn}
\end{equation}
which contains no $\tau$ dependence. When $m\ne n$, the off-diagonal term
takes the following form
\begin{equation}
\begin{split}
  \omega_{mn}+\omega_{nm} ={}&
  \mathcal{P}_{mn}\,e^{-(\Delta E_m+\Delta E_n)t/2}\\
  &\times\cosh\!\left[(E_m{-}E_n)(\tau{-}t/2)\right] ,
\end{split}
  \label{eq:arch}
\end{equation}
where $\mathcal{P}_{mn}\equiv2\zeta_m\zeta_nE_0^2/(\zeta_0^2E_mE_n)$.
Each pair is symmetric under $\tau\to t-\tau$ and is smallest at $\tau=t/2$; its midpoint-to-edge ratio is $\mathrm{sech}[(E_m-E_n)t/2]$, which fixes the shape of the arch, and its amplitude decays as $e^{-(\Delta E_m+\Delta E_n)t/2}$. The contamination is therefore a $\tau$-flat part, whose amplitude is the difference of the diagonal ratios at $z$ and $0$, and an arched part, whose amplitude is the corresponding difference of off-diagonal ratios; both amplitudes are observables.

\subsection{Non-forward kinematics and GPDs}
\label{sec:nonbreit}

The derivation above also applies to GPD matrix elements with non-forward kinematics. 
In this case, $D_{00}$ yields the elastic form factor instead of the local charge.

Let the initial and final states carry momenta $\vec{P}_i$ and $\vec{P}_f\neq\vec{P}_i$, and define
\begin{equation}
  \vec{P}=\tfrac12(\vec{P}_i+\vec{P}_f) ,\qquad
  \vec{\Delta}=\vec{P}_f-\vec{P}_i ,\qquad
  \xi=-\frac{\Delta_z}{P_{i,z}+P_{f,z}} ,
  \label{eq:kinematics}
\end{equation}
with $\Delta^2\equiv\vec{\Delta}^2$ the momentum transfer. The spectra differ, and $C_3^{f\leftarrow i}$ is Eq.~\eqref{eq:c3} with the substitutions $O_{mn}\to O^{fi}_{mn}$, $\zeta_m\zeta_n\to \zeta^f_m\zeta^i_n$, $E_m\to E^f_m$, $E_n\to E^i_n$, where $O^{fi}_{mn}(z)\equiv\langle m^f|\mathcal{O}^{\Gamma}(z)|n^i\rangle$, $D^{fi}_{mn}\equiv O^{fi}_{mn}(0)=\langle m^f|\mathcal{O}^{\Gamma}(0)|n^i\rangle$, and $\zeta^{i,f}_n$ are the overlaps of the initial and final interpolating fields with the $n$-th state; the state at the sink ($m$) carries the final momentum and the state at the source ($n$) the initial one. The object of interest is $h_B^{fi}(z)\equiv O^{fi}_{00}(z)$.

It is worth pointing out that the naive generalization of the conventional ratio to $C_3^{f\leftarrow i}/C_2^f$ is not a consistent estimator:
\begin{equation}
  \frac{C_3^{f\leftarrow i}}{C_2^f}
  \xrightarrow[t\to\infty]{}\,
  \frac{O^{fi}_{00}(z)}{2E^i_0}\frac{\zeta^i_0}{\zeta^f_0}
  \,e^{(E^f_0-E^i_0)\tau} ,
  \label{eq:naive-bad}
\end{equation}
which carries an uncancelled overlap ratio and a factor growing with $\tau$; dividing by $C_2^i$ only exchanges $\tau$ for $t-\tau$. The standard remedy is (the subscript NB denotes the non-Breit frame)
\begin{equation}
  R_{\rm NB}(t,\tau,z) = \frac{C_3^{f\leftarrow i}(t,\tau,z)}{C_2^f(t)}
  \left[
  \frac{C_2^i(t-\tau)C_2^f(\tau)C_2^f(t)}
  {C_2^f(t-\tau)C_2^i(\tau)C_2^i(t)}
  \right]^{1/2} ,
  \label{eq:rnb}
\end{equation}
where $C_2^i$ and $C_2^f$ are the two-point functions of the initial and final interpolating fields, with plateau
\begin{equation}
  R_{\rm NB} \xrightarrow[t\to\infty]{}\,
  {\rm sign}\!\left(\zeta^i_0\zeta^f_0\right)\frac{O^{fi}_{00}(z)}{2\sqrt{E^i_0E^f_0}} .
  \label{eq:rnb-plateau}
\end{equation}
It is awkward in practice: the energy denominator depends on the spectral convention, the construction needs six correlators, a sign ambiguity enters through ${\rm sign}(\zeta^i_0\zeta^f_0)$, and the square root mixes their fluctuations.

The self-ratio strategy avoids this completely. For GPDs, the three-point correlators in Eq.~\eqref{eq:selfratio} simply become $C_3\to C_3^{f\leftarrow i}$, and 
\begin{equation}
R_{\rm self}^{\rm NB}(t,\tau,z)=\frac{C_3^{f\leftarrow i}(t,\tau,z)}{C_3^{f\leftarrow i}(t,\tau,0)}.   
\end{equation}
The cancellation is similar to that in the forward case. Therefore, the non-forward kinematics does not introduce any extra complexity in the self-ratio method. The exact ratio is Eq.~\eqref{eq:selfratio-exact} with $O_{mn}\to O^{fi}_{mn}$ and $D_{mn}\to D^{fi}_{mn}$; the plateau is
\begin{equation}
  R_{\rm self}^{\rm NB}(t,\tau,z)
  \xrightarrow[t\to\infty]{}\,
  \frac{O^{fi}_{00}(z)}{D^{fi}_{00}} ,
  \label{eq:selfratio-nb-plateau}
\end{equation}
and no energy denominator, sign ambiguity or square root appears here. The kinematic prefactor is supplied only at the final stage, when the bare matrix element is converted into a GPD. The first-order residual is that of Eq.~\eqref{eq:selfratio-firstorder} with $O^{fi}_{mn}/O^{fi}_{00}-D^{fi}_{mn}/D^{fi}_{00}$, which vanishes identically at $z=0$ by construction.

\subsection{Transverse-momentum-dependent distributions}
\label{sec:tmd}

The construction carries over when the Wilson line leaves the $\hat{z}$ axis. For a TMD distribution the operator is Eq.~\eqref{eq:op} with the straight line replaced by a staple-shaped contour,
$\mathcal{O}^{\Gamma}(z,b)=\bar{q}(0)\Gamma W_{\rm staple}[(0),(z\hat{z}+b_\perp)]q(z\hat{z}+b_\perp)$,
running from $0$ to $\mathcal{L}\hat{z}$, then to $L\hat{z}+b_\perp$, then to $z\hat{z}+b_\perp$; $\mathcal{L}$ is the arm length, $b_\perp$ the transverse separation conjugate to the parton transverse momentum, with $b\equiv|b_\perp|$, and the contour is purely spatial so that the operator sits on a single time slice. The self-ratio can be constructed as
\begin{equation}
  R_{\rm self}(t,\tau;z,b) \equiv \frac{C_3(t,\tau;z,b)}{C_3(t,\tau;0,b)} ,
  \label{eq:tmd-ratio}
\end{equation}
with numerator and denominator measured with the same source, sink, $t$ and $\tau$. 
One may also try the following double ratio
\begin{equation}
  R^D_{\rm self}(t,\tau;z,b) \equiv \frac{C_3(t,\tau;z,b)}{C_3(t,\tau;0,b)}\frac{C_3(t,\tau;0,0)}{C_3(t,\tau;z,0)}  ,
  \label{eq:tmd-dbratio}
\end{equation}
which is expected to yield a cleaner signal, but the extraction of TMDs becomes more cumbersome due to the appearance of the $z$-dependent factor ${C_3(t,\tau;z,0)}$.

The spectral decomposition, first-order expansion and exact ratio are those of Eqs.~\eqref{eq:c3}, \eqref{eq:Sigma}, \eqref{eq:selfratio-firstorder} and~\eqref{eq:selfratio-ratio}, with $O_{mn}(z)\to O_{mn}(z,b)\equiv\langle m|\mathcal{O}^{\Gamma}(z,b)|n\rangle$, $D_{mn}\to D_{mn}(b)\equiv\langle m|\mathcal{O}^{\Gamma}(0,b)|n\rangle$ and $h_B(z)\to h_B(z,b)\equiv O_{00}(z,b)$; at $b=0$ these reduce to $O_{mn}$, $D_{mn}$ and $h_B(z)$. 

\subsection{Methodological assessment}
\label{sec:assessment}

The self-ratio removes the whole two-point term $\Pi(t)$ and the explicit $1/C_2$ normalization. The overlap ratios $\zeta_n/\zeta_0$ survive in the weights $\omega_{mn}$, but they are shared by every separation and constrained by $C_2(t)$ and $C_3(t,\tau,0)$, so the two-point modeling uncertainty is reduced. It is therefore cheaper than the GEVP or Lanczos, which need new bases or inversions; and less redundant than the summation, Feynman--Hellmann and Bayesian fits, which reweigh the same $C_3/C_2$ data instead of normalizing it.

\section{Case Study}
\label{sec:case}

In this section, we illustrate the strategy developed in Sec.~\ref{sec:theory} through applications to four different lattice observables: the isovector quark transversity PDF of the nucleon, the unpolarized gluon PDF of the nucleon, the quark GPD of the pion, and the nucleon Boer-Mulders (BM) function. For the transversity PDF, we reanalyze the published lattice data of Ref.~\cite{LatticeParton:2022xsd} on the CLS ensemble N203; for the gluon PDF, we use the gluon data of Ref.~\cite{Chen:2025gluon} on the CLQCD ensemble C32P23; the pion GPD is computed on the CLQCD ensemble C24P29 with gauge configurations of Refs.~\cite{Hu:2023qmc,CLQCD:2024tad}; the nucleon BM analysis is based on the published lattice data obtained on the CLS ensemble X650~\cite{LPC:2025spt}. We focus on the bare matrix element fitting only, with the purpose of demonstrating the applicability of the self-ratio to a representative set of observables in the lattice parton program.

These ensembles are collected in Table~\ref{tab:ensembles}. All are $N_f=2+1$. N203 and X650 are CLS ensembles generated with the tree-level Symanzik improved gauge action and 2-HEX smeared clover fermions; C32P23 and C24P29 are CLQCD ensembles generated with the tadpole-improved tree-level Symanzik gauge action (TITLS) and the tadpole-improved tree-level clover fermion action (TITLC). The lattice spacings and pion masses are those quoted in Refs.~\cite{LatticeParton:2022xsd,LPC:2025spt,Hu:2023qmc,CLQCD:2024tad}.

\begin{table*}[t]
  \centering
  \small
  \caption{Ensembles used in this section~\cite{LatticeParton:2022xsd,LPC:2025spt,Hu:2023qmc,CLQCD:2024tad,Chen:2025gluon}. All are $N_f=2+1$, with $L_s^3\times L_t$ the spatial and temporal extents and $m_\pi$ the pion mass. 2-HEX denotes two iterations of HEX (hypercubic blocking) smearing of the gauge links in the clover action; TITLS and TITLC denote the tadpole-improved tree-level Symanzik gauge action and the tadpole-improved tree-level clover fermion action.}
  \label{tab:ensembles}
  \begin{tabular*}{\textwidth}{@{\extracolsep{\fill}}llccc@{}}
    \hline\hline
    Ensemble & Sea fermion / gauge & $L_s^3\times L_t$ & $a$ (fm) & $m_\pi$ (MeV) \\
    \hline
    N203 (CLS) & 2-HEX clover / Symanzik & $48^3\times128$ & 0.064 & 348 \\
    X650 (CLS) & 2-HEX clover / Symanzik & $48^3\times48$ & 0.098 & 338 \\
    C32P23 (CLQCD) & TITLC / TITLS & $32^3\times64$ & 0.105 & 228 \\
    C24P29 (CLQCD) & TITLC / TITLS & $24^3\times72$ & 0.105 & 292 \\
    \hline\hline
  \end{tabular*}
\end{table*}

In all figures of this section the vertical axis is the normalized matrix element
\begin{equation}
  h_B^{\rm norm}(z,b) \equiv \frac{O_{00}(z,b)}{D_{00}(b)} ,
  \label{eq:hbnorm}
\end{equation}
with $b=0$ for the first three cases. The plotted points are the fitted plateaus of the respective estimators divided by their value at $z=0$: for the self-ratio this is the plateau of Eq.~\eqref{eq:selfratio-ratio}, which equals $\hat h_B(z)$ of Eq.~\eqref{eq:hatdef} up to the residual of Eq.~\eqref{eq:selfratio-ratio}, and for the conventional ratio and the Feynman--Hellmann estimator it is defined analogously. All are unity at $z=0$ by construction.

\subsection{Nucleon quark transversity PDF}
\label{sec:case-transversity}

The isovector quark transversity distribution $\delta u(x)-\delta d(x)$ of the nucleon describes the correlation between the transverse spin of the nucleon and that of its quarks, and is the chiral-odd partner of the unpolarized and helicity distributions. At leading twist it is defined from the light-cone correlator with Dirac structure $\gamma^+\gamma^\perp\gamma_5$~\cite{LatticeParton:2022xsd}. On the lattice we use the nonlocal operator of Eq.~\eqref{eq:op} with the Dirac structure appropriate to a transversely polarized nucleon,
\begin{equation}
  h_B^{T}(z;P_z) = \langle P,S_\perp|\,\bar q(z)\,
  \gamma^t\gamma^\perp\gamma_5\,
  W(z,0)\,q(0)\,|P,S_\perp\rangle ,
  \label{eq:trans-op}
\end{equation}
where $S_\perp$ is a transverse polarization, $\perp=x$ or $y$, and the flavor combination $\delta u-\delta d$ is understood, so that disconnected contributions cancel. The Euclidean matrix $\gamma^t\gamma^\perp\gamma_5$ is proportional to $\sigma^{t\perp}\gamma_5$ and is the quasiparton analogue of the light-cone $\gamma^+\gamma^\perp\gamma_5$. At $z=0$ the matrix element reduces to the tensor charge $g_T$ of the nucleon, which provides the reference $h_B^{T}(0)$.

\begin{figure}[htbp]
  \centering
  \includegraphics[width=0.45\textwidth]{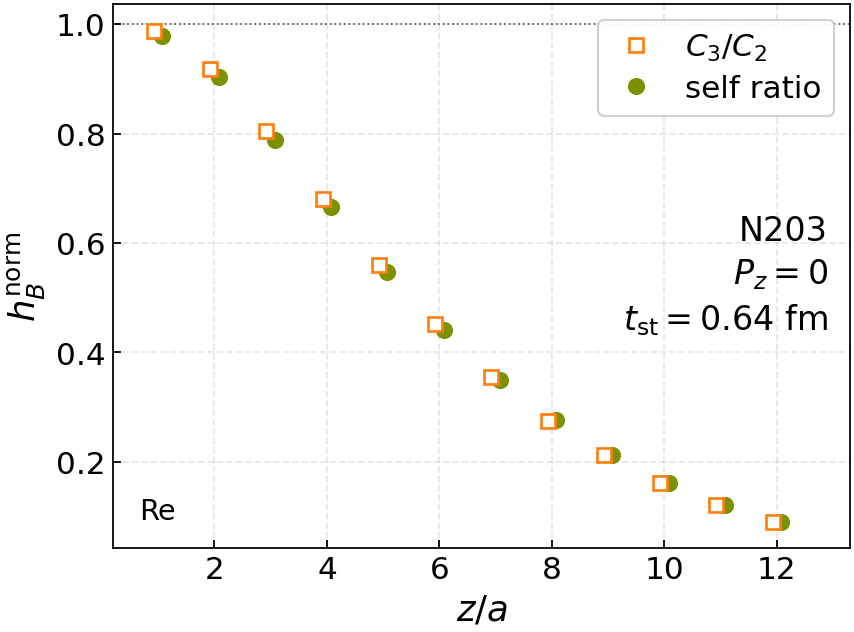}
  \caption{Normalized isovector quark transversity matrix element $h_B^{\rm norm}(z)$ of the nucleon on ensemble N203 at $P_z=0$. Orange/green circles represent the results extracted from the conventional ratio $C_3/C_2$ and the self-ratio, respectively. The two estimators agree within errors.}
  \label{fig:trans-pz0}
\end{figure}

We reanalyse the published data of Ref.~\cite{LatticeParton:2022xsd}, in which the ground-state matrix element was extracted from the ratio $C_3/C_2$ with a two-state fit on ensemble N203. The zero-momentum data provide a closure test of the self-ratio: at $P_z=0$ the matrix element is real and the conventional ratio is under good control, so the two estimators should return the same $h_B^{\rm norm}(z,b)$. Figure~\ref{fig:trans-pz0} shows that they do, point by point, over the whole range $0\le z\le 12a$. This is the expected behaviour when the excited-state residual is small. 

\begin{figure}[htbp]
  \centering
  \includegraphics[width=0.45\textwidth]{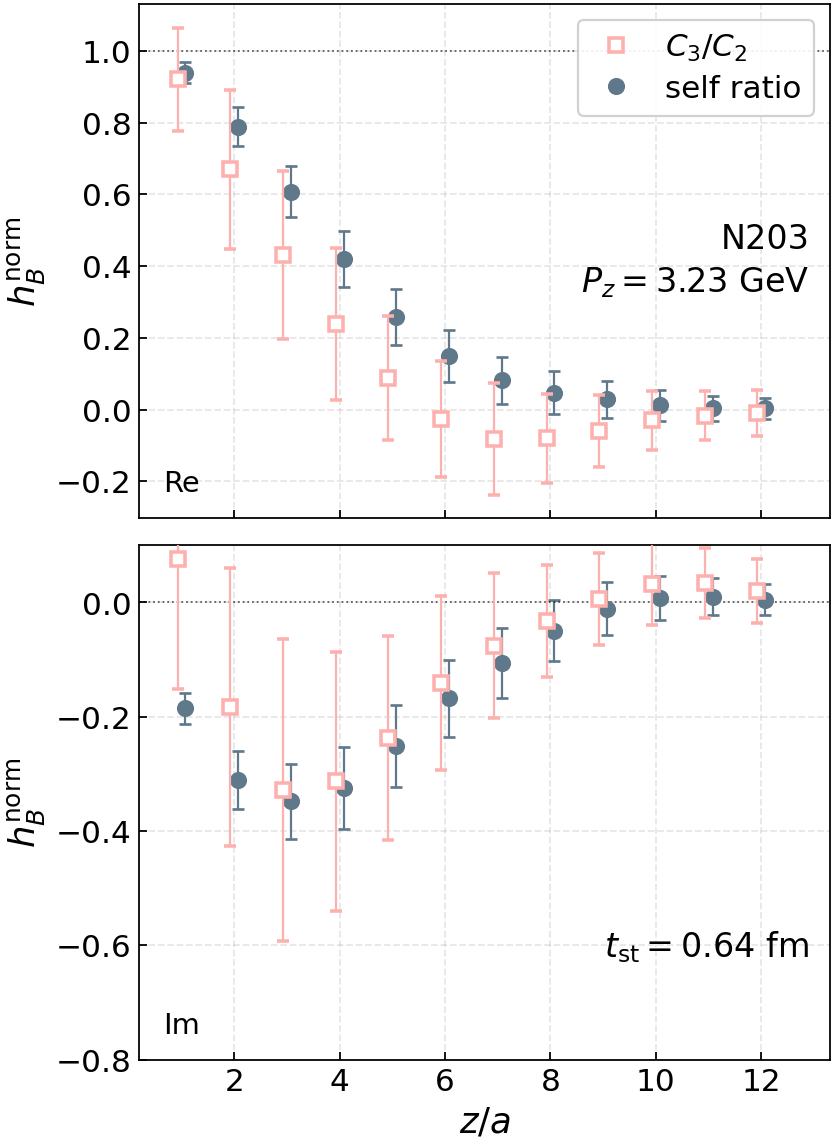}
  \caption{Same as Fig.~\ref{fig:trans-pz0} but at $P_z=3.23$ GeV, showing the real (upper) and imaginary (lower) parts. The conventional ratio loses signal beyond $z\simeq5a$, while the self-ratio remains controlled up to $z\simeq12a$.}
  \label{fig:trans-pz8}
\end{figure}

The advantage of self-ratio method appears at large momentum. Figure~\ref{fig:trans-pz8} shows the same comparison at $P_z=3.23$ GeV. The conventional ratio loses its signal beyond $z\simeq5a$: its error bars grow rapidly and the central values scatter, because the normalization by $C_2$ divides a small three-point function by an exponentially decaying two-point function, so the statistical noise of $C_2$ is amplified into the ratio and the usable fit window shrinks. The self-ratio remains controlled up to $z\simeq12a$ with errors several times smaller, because the common exponential and overlap factors cancel between the two three-point functions and the ratio is formed from quantities with correlated fluctuations. The imaginary part, which is nonzero at finite momentum, shows the same pattern. This is the regime that dominates the LaMET systematics, and the self-ratio extends the reach in $z$, and hence in $x$, at fixed statistics.

\subsection{Nucleon unpolarized gluon PDF}
\label{sec:case-gluonpdf}

For the unpolarized gluon PDF, the analysis of Ref.~\cite{Chen:2025gluon} uses the following quasi-LF correlator,
\begin{equation}
\begin{split}
  h_B^{g}(z;P_z) ={}& \langle P_z|\,\big[M_{tx;tx}(z) + M_{ty;ty}(z)\\
  &- 2M_{xy;xy}(z)\big]\,|P_z\rangle ,
\end{split}
  \label{eq:gluon-op}
\end{equation}
with
\begin{equation}
\begin{split}
  M_{\mu\lambda;\nu\rho}(z) ={}& \sum_{\vec{x}}{\rm Tr}\big[F_{\mu\lambda}(\vec{x}+z\hat{z})\\
  &\times U(\vec{x}+z\hat{z},\vec{x})\,F_{\nu\rho}(\vec{x})\,U(\vec{x},\vec{x}+z\hat{z})\big] ,
\end{split}
  \label{eq:gluon-m}
\end{equation}
where $F_{\mu\nu}$ is the field-strength tensor, $U$ is the Wilson line along $\hat{z}$ in the fundamental representation, and the trace is over color. The gluon operator used here has been shown to be multiplicatively renormalizable~\cite{Zhang:2018diq}. The matrix element is real and even in $z$; the normalized bare matrix element of Ref.~\cite{Chen:2025gluon} differs from Eq.~\eqref{eq:gluon-op} only by the $z=0$ normalization, i.e. it is $h_B^{\rm norm}(z,b{=}0)$ of Eq.~\eqref{eq:hbnorm}.

\begin{figure}[htbp]
  \centering
  \includegraphics[width=0.45\textwidth]{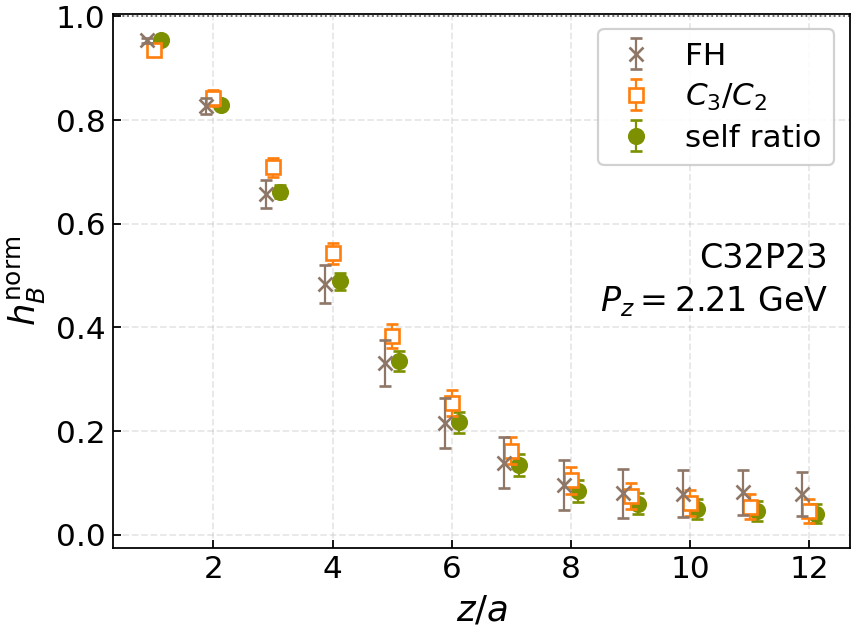}
  \caption{Normalized unpolarized gluon matrix element $h_B^{\rm norm}(z,b{=}0)$ on ensemble C32P23 at $P_z=2.21$ GeV from the Feynman--Hellmann method (blue squares), the conventional ratio (orange circles), and the self-ratio (green circles).}
  \label{fig:gluon}
\end{figure}

We consider the matrix element on ensemble C32P23 at $P_z=2.21$ GeV; the data are those of Ref.~\cite{Chen:2025gluon}. Figure~\ref{fig:gluon} compares three determinations of $h_B^{\rm norm}(z,b)$: the Feynman--Hellmann method (squares), the conventional ratio $C_3/C_2$ (filled circles), and the self-ratio (open circles). All three agree within uncertainties, with the conventional ratio lying somewhat above the other two at intermediate separations and the self-ratio reaching the same plateau with comparable or smaller errors. The agreement is a check of estimator independence, because the three estimators have different sensitivity to the excited-state contamination and to the two-point normalization.

\subsection{Pion quark GPD}
\label{sec:case-piongpd}

The pion quark GPD $H^\pi(x,\xi,\Delta^2)$ correlates the longitudinal momentum fraction with the transverse position of quarks in the pion and provides a tomography of the Goldstone boson of chiral symmetry breaking~\cite{Chen:2019lcm,Lin:2023gxz,Ding:2024saz}. It is accessed through the non-forward matrix element of Eq.~\eqref{eq:op} with $\Gamma=\gamma_t$, $h_B^{\pi}(z,\xi,\Delta^2)=\langle \pi(P_f)|\bar u(0)\gamma_t W(0,z\hat{z})u(z\hat{z})|\pi(P_i)\rangle$, where $u$ is the up-quark field, $\vec{P}_f-\vec{P}_i=\vec{\Delta}$ and $\xi$ the skewness of Eq.~\eqref{eq:kinematics}; this is the $i\to f$ matrix element $h_B^{fi}(z)$ of Sec.~\ref{sec:nonbreit}, and its spectral decomposition is the non-forward form of Eq.~\eqref{eq:c3}. At $z=0$ it reduces to the electromagnetic form factor $F_\pi(\Delta^2)$ up to the standard normalization. We compute it on ensemble C24P29 at $\xi=-0.50$~\cite{Zhao:toappear}, with the gauge configurations of Refs.~\cite{Hu:2023qmc,CLQCD:2024tad}. As Sec.~\ref{sec:nonbreit} shows, the self-ratio needs neither the six correlators of Eq.~\eqref{eq:rnb} nor its sign and square-root ambiguities. 

\begin{figure}[htbp]
  \centering
  \includegraphics[width=0.45\textwidth]{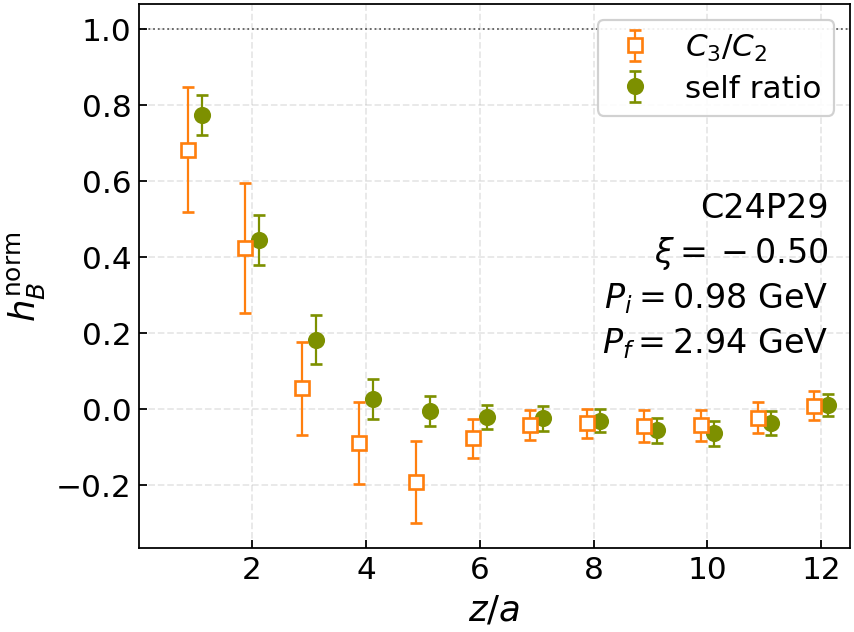}
  \caption{Real part of the normalized pion quark GPD matrix element $h_B^{\rm norm}(z,b{=}0)$ on ensemble C24P29 at $\xi=-0.50$, with $P_i=0.98$ GeV and $P_f=2.94$ GeV, from the time-symmetrized ratio $R_{\rm NB}$ of Eq.~\eqref{eq:rnb} (orange circles) and the self-ratio (green circles). The conventional ratio dips and changes sign at intermediate $z$, while the self-ratio remains close to zero. The imaginary part is not shown because its signal is poor at the present statistics.}
  \label{fig:gpd}
\end{figure}

Figure~\ref{fig:gpd} shows the resulting $h_B^{\rm norm}(z,b)$ from the conventional ratio and from the self-ratio. The conventional ratio dips at intermediate separations, around $z\simeq3$--$6a$, and changes sign, while the self-ratio stays close to zero over the same range. The dip is absent from the self-ratio, which indicates that it is carried by the contamination cancelled in Eq.~\eqref{eq:common-mode} rather than by the matrix element itself: in the conventional ratio the two-point normalization and the excited-state terms enter with full strength and are largest at intermediate $z$, where the two-point functions of Eq.~\eqref{eq:rnb} are evaluated at different time arguments. The self-ratio cancels that common structure, so the extracted matrix element is smooth and the plateau can be fitted over a wider range. This is the practical advantage of the method for GPDs anticipated in Sec.~\ref{sec:nonbreit}.

\subsection{Nucleon Boer-Mulders function}
\label{sec:case-bm}

The Boer-Mulders function $h_1^\perp(x,k_T^2)$ is the leading-twist TMD that correlates the transverse spin of a quark with its transverse momentum in an unpolarized hadron~\cite{Boer:1997nt,Mulders:1995dh}. It is naive time-reversal odd, so it requires a staple-shaped Wilson line and flips sign between the SIDIS and Drell-Yan processes. The corresponding quasi-LF correlator follows from Ref.~\cite{LPC:2025spt},
\begin{equation}
\begin{split}
  h_B^{\rm BM}(z,b_\perp;P_z) ={}& \langle P|\,\bar q(b_\perp)\,
  i\sigma^{yt}\gamma_5\\
  &\times W_{\sqsubset}(z,b_\perp)\,q(z\hat{z})\,|P\rangle ,
\end{split}
  \label{eq:bm-op}
\end{equation}
where $|P\rangle$ is an unpolarized nucleon, $y$ is the transverse index of the quark spin, and $W_{\sqsubset}(z,b_\perp)$ is the staple-shaped gauge link running from $z\hat{z}$ to $(z+\mathcal{L})\hat{z}$, then to $(z+\mathcal{L})\hat{z}+b_\perp$, then to $b_\perp$; the trace over color and spin is understood. 

\begin{figure}[htbp]
  \centering
  \includegraphics[width=0.45\textwidth]{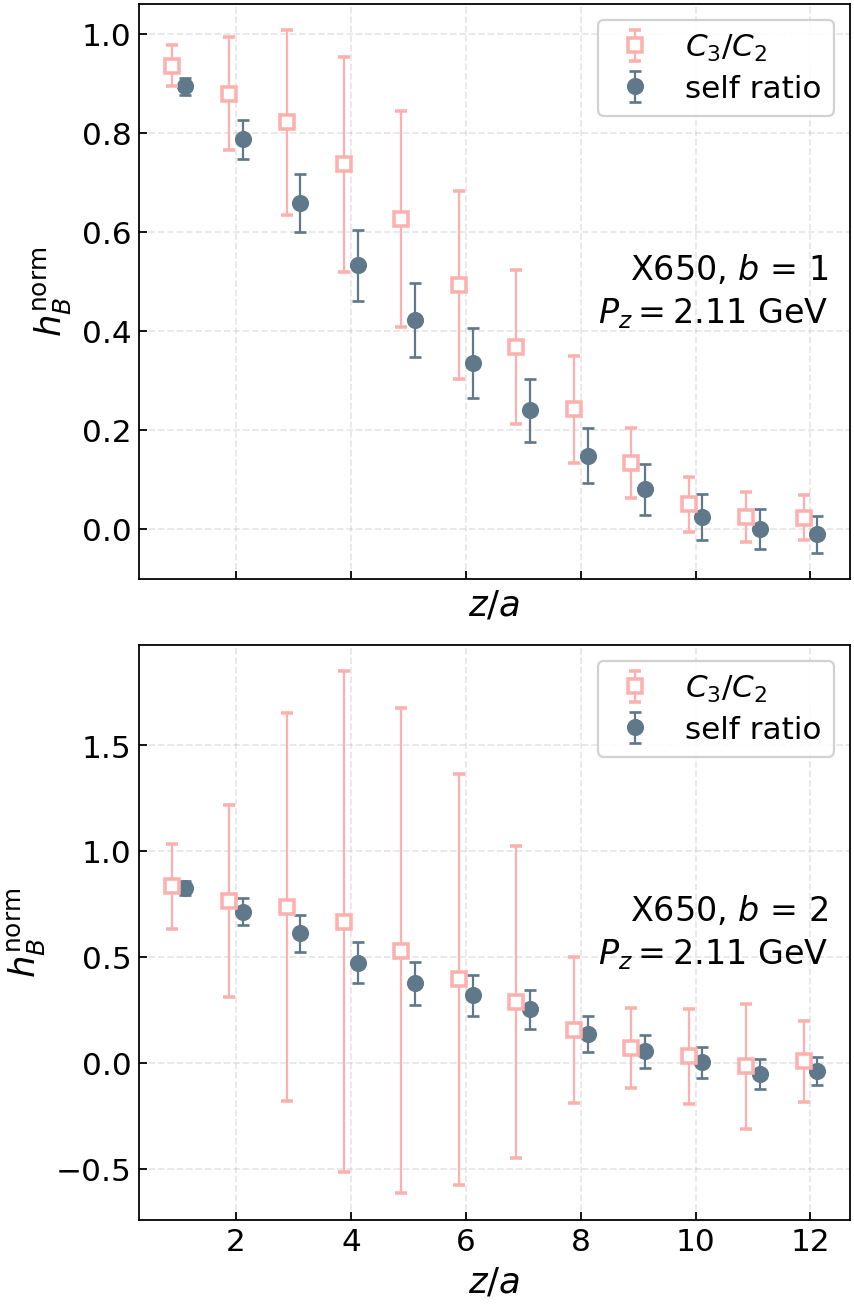}
  \caption{Normalized Boer-Mulders matrix element $h_B^{\rm norm}(z,b)$ on ensemble X650 at $P_z=2.11$ GeV for $b/a=1$ (upper) and $b/a=2$ (lower), from the conventional ratio (orange circles) and the self-ratio (green circles).}
  \label{fig:bm}
\end{figure}

We refit the matrix element on ensemble X650~\cite{LPC:2025spt} at $P_z=2.11$ GeV for $b/a=1$ and $2$. Figure~\ref{fig:bm} shows the resulting $h_B^{\rm norm}(z,b)$ from the conventional ratio and from the self-ratio in Eq.~(\ref{eq:tmd-ratio}). The double ratio in Eq.~(\ref{eq:tmd-dbratio}) is not directly applicable here as the matrix element vanishes at $b=0$ in the present case. As can be seen from the Figure, both estimators decrease with $z$, as expected at fixed transverse separation, and their overall magnitude decreases from $b/a=1$ to $b/a=2$, the expected decay with transverse separation. The result from the self-ratio method agrees with that from the conventional ratio within errors, but has visibly smaller errors in the entire range of $z$. 

\section{Summary and outlook}
\label{sec:summary}

We have proposed a self-ratio fitting strategy for the nonlocal correlators underlying lattice calculations of parton physics, and derived its spectral decomposition in both forward and nonforward kinematics. Since the numerator and denominator of the self-ratio share the same overlap and energy factors, it allows excited-state contributions with similar relative weights to cancel and therefore substantially reduces the excited-state contamination relative to the conventional ratio of three-point to two-point correlators.

We have demonstrated the method for four representative examples: the quark transversity PDF and unpolarized gluon PDF of the nucleon, the quark GPD of the pion, and the Boer–Mulders function of the nucleon. Across all four applications, the self-ratio yields results consistent with conventional determinations where the latter are reliable, while achieving smaller statistical uncertainties. Within the ranges studied, it also remains robust at large hadron momenta and large separations, where the conventional ratio becomes increasingly noisy or loses signal. Importantly, the method can generally be implemented using the correlator data already available in standard calculations, without requiring additional measurements.

Although our demonstrations focus on quasi-LF correlators, the underlying strategy is not restricted to this setting. The same principle can be naturally extended to other nonlocal correlators, including current–current correlators and those probing multiparton physics.

These results suggest that the self-ratio offers a practical means of improving the precision and extending the kinematic reach of lattice parton physics calculations, and we plan to incorporate it into future analyses. A potential improvement of this method is to quantify the remaining excited-state effects through a systematic study of the $z$-dependent residual contributions, guided by the spectral decomposition. In addition, it would be interesting to investigate how increasing the number of states retained in the spectral parametrization affects the extraction of ground-state matrix elements.

It will also be valuable to assess how far the accessible hadron-momentum range can be extended using the self-ratio at fixed computational cost, while maintaining control over excited-state contamination and other systematic effects. Such a study would clarify whether the improved statistical precision can be translated into better control of finite-momentum corrections and, ultimately, more reliable determinations of parton physics.

\begin{acknowledgments}
We thank Chen Chen, Lingquan Ma, Xiaonu Xiong, Fei Yao, Chunhua Zeng, and the LPC collaboration for providing the lattice data used in our numerical tests, and thank Peng Sun, Jinxin Tan and other LPC members for valuable discussions. The work is supported in part by National Natural Science Foundation of China under grant No. 12375080, the Ministry of Science and Technology of China under Grant No. 2024YFA1611004, and by CUHK-Shenzhen under grant No. UDF01002851.
\end{acknowledgments}

\bibliography{main}

\begin{thebibliography}{46}%
\makeatletter
\providecommand \@ifxundefined [1]{%
 \@ifx{#1\undefined}
}%
\providecommand \@ifnum [1]{%
 \ifnum #1\expandafter \@firstoftwo
 \else \expandafter \@secondoftwo
 \fi
}%
\providecommand \@ifx [1]{%
 \ifx #1\expandafter \@firstoftwo
 \else \expandafter \@secondoftwo
 \fi
}%
\providecommand \natexlab [1]{#1}%
\providecommand \enquote  [1]{``#1''}%
\providecommand \bibnamefont  [1]{#1}%
\providecommand \bibfnamefont [1]{#1}%
\providecommand \citenamefont [1]{#1}%
\providecommand \href@noop [0]{\@secondoftwo}%
\providecommand \href [0]{\begingroup \@sanitize@url \@href}%
\providecommand \@href[1]{\@@startlink{#1}\@@href}%
\providecommand \@@href[1]{\endgroup#1\@@endlink}%
\providecommand \@sanitize@url [0]{\catcode `\\12\catcode `\$12\catcode
  `\&12\catcode `\#12\catcode `\^12\catcode `\_12\catcode `\%12\relax}%
\providecommand \@@startlink[1]{}%
\providecommand \@@endlink[0]{}%
\providecommand \url  [0]{\begingroup\@sanitize@url \@url }%
\providecommand \@url [1]{\endgroup\@href {#1}{\urlprefix }}%
\providecommand \urlprefix  [0]{URL }%
\providecommand \Eprint [0]{\href }%
\providecommand \doibase [0]{https://doi.org/}%
\providecommand \selectlanguage [0]{\@gobble}%
\providecommand \bibinfo  [0]{\@secondoftwo}%
\providecommand \bibfield  [0]{\@secondoftwo}%
\providecommand \translation [1]{[#1]}%
\providecommand \BibitemOpen [0]{}%
\providecommand \bibitemStop [0]{}%
\providecommand \bibitemNoStop [0]{.\EOS\space}%
\providecommand \EOS [0]{\spacefactor3000\relax}%
\providecommand \BibitemShut  [1]{\csname bibitem#1\endcsname}%
\let\auto@bib@innerbib\@empty
\bibitem [{\citenamefont {Accardi}\ \emph {et~al.}(2016)\citenamefont {Accardi}
  \emph {et~al.}}]{Accardi:2012qut}%
  \BibitemOpen
  \bibfield  {author} {\bibinfo {author} {\bibfnamefont {A.}~\bibnamefont
  {Accardi}} \emph {et~al.},\ }\bibfield  {title} {\bibinfo {title} {{Electron
  Ion Collider: The Next QCD Frontier}: {Understanding the glue that binds us
  all}},\ }\href {https://doi.org/10.1140/epja/i2016-16268-9} {\bibfield
  {journal} {\bibinfo  {journal} {Eur. Phys. J. A}\ }\textbf {\bibinfo {volume}
  {52}},\ \bibinfo {pages} {268} (\bibinfo {year} {2016})},\ \Eprint
  {https://arxiv.org/abs/1212.1701} {arXiv:1212.1701 [nucl-ex]} \BibitemShut
  {NoStop}%
\bibitem [{\citenamefont {Abdul~Khalek}\ \emph {et~al.}(2022)\citenamefont
  {Abdul~Khalek} \emph {et~al.}}]{AbdulKhalek:2021gbh}%
  \BibitemOpen
  \bibfield  {author} {\bibinfo {author} {\bibfnamefont {R.}~\bibnamefont
  {Abdul~Khalek}} \emph {et~al.},\ }\bibfield  {title} {\bibinfo {title}
  {{Science Requirements and Detector Concepts for the Electron-Ion Collider}:
  {EIC Yellow Report}},\ }\href
  {https://doi.org/10.1016/j.nuclphysa.2022.122447} {\bibfield  {journal}
  {\bibinfo  {journal} {Nucl. Phys. A}\ }\textbf {\bibinfo {volume} {1026}},\
  \bibinfo {pages} {122447} (\bibinfo {year} {2022})},\ \Eprint
  {https://arxiv.org/abs/2103.05419} {arXiv:2103.05419 [physics.ins-det]}
  \BibitemShut {NoStop}%
\bibitem [{\citenamefont {Anderle}\ \emph {et~al.}(2021)\citenamefont {Anderle}
  \emph {et~al.}}]{Anderle:2021wcy}%
  \BibitemOpen
  \bibfield  {author} {\bibinfo {author} {\bibfnamefont {D.~P.}\ \bibnamefont
  {Anderle}} \emph {et~al.},\ }\bibfield  {title} {\bibinfo {title}
  {{Electron-ion collider in China}},\ }\href
  {https://doi.org/10.1007/s11467-021-1062-0} {\bibfield  {journal} {\bibinfo
  {journal} {Front. Phys. (Beijing)}\ }\textbf {\bibinfo {volume} {16}},\
  \bibinfo {pages} {64701} (\bibinfo {year} {2021})},\ \Eprint
  {https://arxiv.org/abs/2102.09222} {arXiv:2102.09222 [nucl-ex]} \BibitemShut
  {NoStop}%
\bibitem [{\citenamefont {Martinelli}\ \emph {et~al.}(1995)\citenamefont
  {Martinelli}, \citenamefont {Pittori}, \citenamefont {Sachrajda},
  \citenamefont {Testa},\ and\ \citenamefont {Vladikas}}]{Martinelli:1994ty}%
  \BibitemOpen
  \bibfield  {author} {\bibinfo {author} {\bibfnamefont {G.}~\bibnamefont
  {Martinelli}}, \bibinfo {author} {\bibfnamefont {C.}~\bibnamefont {Pittori}},
  \bibinfo {author} {\bibfnamefont {C.~T.}\ \bibnamefont {Sachrajda}}, \bibinfo
  {author} {\bibfnamefont {M.}~\bibnamefont {Testa}},\ and\ \bibinfo {author}
  {\bibfnamefont {A.}~\bibnamefont {Vladikas}},\ }\bibfield  {title} {\bibinfo
  {title} {{A general method for non-perturbative renormalization of lattice
  operators}},\ }\href {https://doi.org/10.1016/0550-3213(95)00126-D}
  {\bibfield  {journal} {\bibinfo  {journal} {Nucl. Phys. B}\ }\textbf
  {\bibinfo {volume} {445}},\ \bibinfo {pages} {81} (\bibinfo {year} {1995})},\
  \Eprint {https://arxiv.org/abs/hep-lat/9411010} {arXiv:hep-lat/9411010}
  \BibitemShut {NoStop}%
\bibitem [{\citenamefont {Detmold}\ and\ \citenamefont
  {Lin}(2006)}]{Detmold:2005gg}%
  \BibitemOpen
  \bibfield  {author} {\bibinfo {author} {\bibfnamefont {W.}~\bibnamefont
  {Detmold}}\ and\ \bibinfo {author} {\bibfnamefont {C.~J.~D.}\ \bibnamefont
  {Lin}},\ }\bibfield  {title} {\bibinfo {title} {{Deep-inelastic scattering
  and the operator product expansion in lattice QCD}},\ }\href
  {https://doi.org/10.1103/PhysRevD.73.014501} {\bibfield  {journal} {\bibinfo
  {journal} {Phys. Rev. D}\ }\textbf {\bibinfo {volume} {73}},\ \bibinfo
  {pages} {014501} (\bibinfo {year} {2006})},\ \Eprint
  {https://arxiv.org/abs/hep-lat/0507007} {arXiv:hep-lat/0507007} \BibitemShut
  {NoStop}%
\bibitem [{\citenamefont {Liu}\ and\ \citenamefont {Dong}(1994)}]{Liu:1993cv}%
  \BibitemOpen
  \bibfield  {author} {\bibinfo {author} {\bibfnamefont {K.-F.}\ \bibnamefont
  {Liu}}\ and\ \bibinfo {author} {\bibfnamefont {S.-J.}\ \bibnamefont {Dong}},\
  }\bibfield  {title} {\bibinfo {title} {{Origin of difference between anti-d
  and anti-u partons in the nucleon}},\ }\href
  {https://doi.org/10.1103/PhysRevLett.72.1790} {\bibfield  {journal} {\bibinfo
   {journal} {Phys. Rev. Lett.}\ }\textbf {\bibinfo {volume} {72}},\ \bibinfo
  {pages} {1790} (\bibinfo {year} {1994})},\ \Eprint
  {https://arxiv.org/abs/hep-ph/9306299} {arXiv:hep-ph/9306299} \BibitemShut
  {NoStop}%
\bibitem [{\citenamefont {Ji}(2013)}]{Ji:2013dva}%
  \BibitemOpen
  \bibfield  {author} {\bibinfo {author} {\bibfnamefont {X.}~\bibnamefont
  {Ji}},\ }\bibfield  {title} {\bibinfo {title} {{Parton Physics on a Euclidean
  Lattice}},\ }\href {https://doi.org/10.1103/PhysRevLett.110.262002}
  {\bibfield  {journal} {\bibinfo  {journal} {Phys. Rev. Lett.}\ }\textbf
  {\bibinfo {volume} {110}},\ \bibinfo {pages} {262002} (\bibinfo {year}
  {2013})},\ \Eprint {https://arxiv.org/abs/1305.1539} {arXiv:1305.1539
  [hep-ph]} \BibitemShut {NoStop}%
\bibitem [{\citenamefont {Ji}(2014)}]{Ji:2014gla}%
  \BibitemOpen
  \bibfield  {author} {\bibinfo {author} {\bibfnamefont {X.}~\bibnamefont
  {Ji}},\ }\bibfield  {title} {\bibinfo {title} {{Parton Physics from
  Large-Momentum Effective Field Theory}},\ }\href
  {https://doi.org/10.1007/s11433-014-5492-3} {\bibfield  {journal} {\bibinfo
  {journal} {Sci. China Phys. Mech. Astron.}\ }\textbf {\bibinfo {volume}
  {57}},\ \bibinfo {pages} {1407} (\bibinfo {year} {2014})},\ \Eprint
  {https://arxiv.org/abs/1404.6680} {arXiv:1404.6680 [hep-ph]} \BibitemShut
  {NoStop}%
\bibitem [{\citenamefont {Ji}\ \emph {et~al.}(2021)\citenamefont {Ji},
  \citenamefont {Liu}, \citenamefont {Liu}, \citenamefont {Zhang},\ and\
  \citenamefont {Zhao}}]{Ji:2020ect}%
  \BibitemOpen
  \bibfield  {author} {\bibinfo {author} {\bibfnamefont {X.}~\bibnamefont
  {Ji}}, \bibinfo {author} {\bibfnamefont {Y.-S.}\ \bibnamefont {Liu}},
  \bibinfo {author} {\bibfnamefont {Y.}~\bibnamefont {Liu}}, \bibinfo {author}
  {\bibfnamefont {J.-H.}\ \bibnamefont {Zhang}},\ and\ \bibinfo {author}
  {\bibfnamefont {Y.}~\bibnamefont {Zhao}},\ }\bibfield  {title} {\bibinfo
  {title} {{Large-momentum effective theory}},\ }\href
  {https://doi.org/10.1103/RevModPhys.93.035005} {\bibfield  {journal}
  {\bibinfo  {journal} {Rev. Mod. Phys.}\ }\textbf {\bibinfo {volume} {93}},\
  \bibinfo {pages} {035005} (\bibinfo {year} {2021})},\ \Eprint
  {https://arxiv.org/abs/2004.03543} {arXiv:2004.03543 [hep-ph]} \BibitemShut
  {NoStop}%
\bibitem [{\citenamefont {Ji}(2024)}]{Ji:2024oka}%
  \BibitemOpen
  \bibfield  {author} {\bibinfo {author} {\bibfnamefont {X.}~\bibnamefont
  {Ji}},\ }\bibfield  {title} {\bibinfo {title} {{Euclidean effective theory
  for partons in the spirit of Steven Weinberg}},\ }\href
  {https://doi.org/10.1016/j.nuclphysb.2024.116670} {\bibfield  {journal}
  {\bibinfo  {journal} {Nucl. Phys. B}\ }\textbf {\bibinfo {volume} {1007}},\
  \bibinfo {pages} {116670} (\bibinfo {year} {2024})},\ \Eprint
  {https://arxiv.org/abs/2408.03378} {arXiv:2408.03378 [hep-ph]} \BibitemShut
  {NoStop}%
\bibitem [{\citenamefont
  {Radyushkin}(2017{\natexlab{a}})}]{Radyushkin:2017cyf}%
  \BibitemOpen
  \bibfield  {author} {\bibinfo {author} {\bibfnamefont {A.~V.}\ \bibnamefont
  {Radyushkin}},\ }\bibfield  {title} {\bibinfo {title} {{Quasi-parton
  distribution functions, momentum distributions, and pseudo-parton
  distribution functions}},\ }\href
  {https://doi.org/10.1103/PhysRevD.96.034025} {\bibfield  {journal} {\bibinfo
  {journal} {Phys. Rev. D}\ }\textbf {\bibinfo {volume} {96}},\ \bibinfo
  {pages} {034025} (\bibinfo {year} {2017}{\natexlab{a}})},\ \Eprint
  {https://arxiv.org/abs/1705.01488} {arXiv:1705.01488 [hep-ph]} \BibitemShut
  {NoStop}%
\bibitem [{\citenamefont
  {Radyushkin}(2017{\natexlab{b}})}]{Radyushkin:2017lvu}%
  \BibitemOpen
  \bibfield  {author} {\bibinfo {author} {\bibfnamefont {A.~V.}\ \bibnamefont
  {Radyushkin}},\ }\bibfield  {title} {\bibinfo {title} {{Nonperturbative
  Evolution of Parton Quasi-Distributions}},\ }\href
  {https://doi.org/10.1016/j.physletb.2017.07.075} {\bibfield  {journal}
  {\bibinfo  {journal} {Phys. Lett. B}\ }\textbf {\bibinfo {volume} {774}},\
  \bibinfo {pages} {200} (\bibinfo {year} {2017}{\natexlab{b}})},\ \Eprint
  {https://arxiv.org/abs/1705.01488} {arXiv:1705.01488 [hep-ph]} \BibitemShut
  {NoStop}%
\bibitem [{\citenamefont {Braun}\ and\ \citenamefont
  {M{\"u}ller}(2008)}]{Braun:2007wv}%
  \BibitemOpen
  \bibfield  {author} {\bibinfo {author} {\bibfnamefont {V.}~\bibnamefont
  {Braun}}\ and\ \bibinfo {author} {\bibfnamefont {D.}~\bibnamefont
  {M{\"u}ller}},\ }\bibfield  {title} {\bibinfo {title} {{Exclusive processes
  in position space and the pion distribution amplitude}},\ }\href
  {https://doi.org/10.1140/epjc/s10052-008-0608-4} {\bibfield  {journal}
  {\bibinfo  {journal} {Eur. Phys. J. C}\ }\textbf {\bibinfo {volume} {55}},\
  \bibinfo {pages} {349} (\bibinfo {year} {2008})},\ \Eprint
  {https://arxiv.org/abs/0709.1348} {arXiv:0709.1348 [hep-ph]} \BibitemShut
  {NoStop}%
\bibitem [{\citenamefont {Ma}\ and\ \citenamefont
  {Qiu}(2018{\natexlab{a}})}]{Ma:2017pxb}%
  \BibitemOpen
  \bibfield  {author} {\bibinfo {author} {\bibfnamefont {Y.-Q.}\ \bibnamefont
  {Ma}}\ and\ \bibinfo {author} {\bibfnamefont {J.-W.}\ \bibnamefont {Qiu}},\
  }\bibfield  {title} {\bibinfo {title} {{Exploring Partonic Structure of
  Hadrons Using ab initio Lattice QCD Calculations}},\ }\href
  {https://doi.org/10.1103/PhysRevLett.120.022003} {\bibfield  {journal}
  {\bibinfo  {journal} {Phys. Rev. Lett.}\ }\textbf {\bibinfo {volume} {120}},\
  \bibinfo {pages} {022003} (\bibinfo {year} {2018}{\natexlab{a}})},\ \Eprint
  {https://arxiv.org/abs/1709.03018} {arXiv:1709.03018 [hep-ph]} \BibitemShut
  {NoStop}%
\bibitem [{\citenamefont {Chambers}\ \emph {et~al.}(2017)\citenamefont
  {Chambers}, \citenamefont {Horsley}, \citenamefont {Nakamura}, \citenamefont
  {Perlt}, \citenamefont {Rakow}, \citenamefont {Schierholz}, \citenamefont
  {Schiller}, \citenamefont {Somfleth}, \citenamefont {Young},\ and\
  \citenamefont {Zanotti}}]{Chambers:2017dov}%
  \BibitemOpen
  \bibfield  {author} {\bibinfo {author} {\bibfnamefont {A.~J.}\ \bibnamefont
  {Chambers}}, \bibinfo {author} {\bibfnamefont {R.}~\bibnamefont {Horsley}},
  \bibinfo {author} {\bibfnamefont {Y.}~\bibnamefont {Nakamura}}, \bibinfo
  {author} {\bibfnamefont {H.}~\bibnamefont {Perlt}}, \bibinfo {author}
  {\bibfnamefont {P.~E.~L.}\ \bibnamefont {Rakow}}, \bibinfo {author}
  {\bibfnamefont {G.}~\bibnamefont {Schierholz}}, \bibinfo {author}
  {\bibfnamefont {A.}~\bibnamefont {Schiller}}, \bibinfo {author}
  {\bibfnamefont {K.}~\bibnamefont {Somfleth}}, \bibinfo {author}
  {\bibfnamefont {R.~D.}\ \bibnamefont {Young}},\ and\ \bibinfo {author}
  {\bibfnamefont {J.~M.}\ \bibnamefont {Zanotti}},\ }\bibfield  {title}
  {\bibinfo {title} {{Nucleon Structure Functions from Operator Product
  Expansion on the Lattice}},\ }\href
  {https://doi.org/10.1103/PhysRevLett.118.242001} {\bibfield  {journal}
  {\bibinfo  {journal} {Phys. Rev. Lett.}\ }\textbf {\bibinfo {volume} {118}},\
  \bibinfo {pages} {242001} (\bibinfo {year} {2017})},\ \Eprint
  {https://arxiv.org/abs/1703.01153} {arXiv:1703.01153 [hep-lat]} \BibitemShut
  {NoStop}%
\bibitem [{\citenamefont {Ji}\ \emph {et~al.}(2018)\citenamefont {Ji},
  \citenamefont {Zhang},\ and\ \citenamefont {Zhao}}]{Ji:2017oey}%
  \BibitemOpen
  \bibfield  {author} {\bibinfo {author} {\bibfnamefont {X.}~\bibnamefont
  {Ji}}, \bibinfo {author} {\bibfnamefont {J.-H.}\ \bibnamefont {Zhang}},\ and\
  \bibinfo {author} {\bibfnamefont {Y.}~\bibnamefont {Zhao}},\ }\bibfield
  {title} {\bibinfo {title} {{Renormalization in Large Momentum Effective
  Theory of Parton Physics}},\ }\href
  {https://doi.org/10.1103/PhysRevLett.120.112001} {\bibfield  {journal}
  {\bibinfo  {journal} {Phys. Rev. Lett.}\ }\textbf {\bibinfo {volume} {120}},\
  \bibinfo {pages} {112001} (\bibinfo {year} {2018})},\ \Eprint
  {https://arxiv.org/abs/1706.08962} {arXiv:1706.08962 [hep-ph]} \BibitemShut
  {NoStop}%
\bibitem [{\citenamefont {Ishikawa}\ \emph {et~al.}(2019)\citenamefont
  {Ishikawa}, \citenamefont {Jin}, \citenamefont {Lin}, \citenamefont
  {Sch{\"a}fer}, \citenamefont {Yang}, \citenamefont {Zhang},\ and\
  \citenamefont {Zhao}}]{Ishikawa:2017iym}%
  \BibitemOpen
  \bibfield  {author} {\bibinfo {author} {\bibfnamefont {T.}~\bibnamefont
  {Ishikawa}}, \bibinfo {author} {\bibfnamefont {L.}~\bibnamefont {Jin}},
  \bibinfo {author} {\bibfnamefont {H.-W.}\ \bibnamefont {Lin}}, \bibinfo
  {author} {\bibfnamefont {A.}~\bibnamefont {Sch{\"a}fer}}, \bibinfo {author}
  {\bibfnamefont {Y.-B.}\ \bibnamefont {Yang}}, \bibinfo {author}
  {\bibfnamefont {J.-H.}\ \bibnamefont {Zhang}},\ and\ \bibinfo {author}
  {\bibfnamefont {Y.}~\bibnamefont {Zhao}},\ }\bibfield  {title} {\bibinfo
  {title} {{Gaussian-weighted parton quasi-distribution (Lattice Parton Physics
  Project (LP$^{3}$))}},\ }\href {https://doi.org/10.1007/s11433-018-9375-1}
  {\bibfield  {journal} {\bibinfo  {journal} {Sci. China Phys. Mech. Astron.}\
  }\textbf {\bibinfo {volume} {62}},\ \bibinfo {pages} {991021} (\bibinfo
  {year} {2019})},\ \Eprint {https://arxiv.org/abs/1711.07858}
  {arXiv:1711.07858 [hep-ph]} \BibitemShut {NoStop}%
\bibitem [{\citenamefont {Green}\ \emph {et~al.}(2018)\citenamefont {Green},
  \citenamefont {Jansen},\ and\ \citenamefont {Steffens}}]{Green:2017xeu}%
  \BibitemOpen
  \bibfield  {author} {\bibinfo {author} {\bibfnamefont {J.}~\bibnamefont
  {Green}}, \bibinfo {author} {\bibfnamefont {K.}~\bibnamefont {Jansen}},\ and\
  \bibinfo {author} {\bibfnamefont {F.}~\bibnamefont {Steffens}},\ }\bibfield
  {title} {\bibinfo {title} {{Nonperturbative Renormalization of Nonlocal Quark
  Bilinears for Parton Quasidistribution Functions on the Lattice Using an
  Auxiliary Field}},\ }\href {https://doi.org/10.1103/PhysRevLett.121.022004}
  {\bibfield  {journal} {\bibinfo  {journal} {Phys. Rev. Lett.}\ }\textbf
  {\bibinfo {volume} {121}},\ \bibinfo {pages} {022004} (\bibinfo {year}
  {2018})},\ \Eprint {https://arxiv.org/abs/1707.07152} {arXiv:1707.07152
  [hep-lat]} \BibitemShut {NoStop}%
\bibitem [{\citenamefont {Liu}\ \emph {et~al.}(2020)\citenamefont {Liu} \emph
  {et~al.}}]{LatticeParton:2018gjr}%
  \BibitemOpen
  \bibfield  {author} {\bibinfo {author} {\bibfnamefont {Y.-S.}\ \bibnamefont
  {Liu}} \emph {et~al.} (\bibinfo {collaboration} {Lattice Parton}),\
  }\bibfield  {title} {\bibinfo {title} {{Unpolarized isovector quark
  distribution function from lattice QCD: A systematic analysis of
  renormalization and matching}},\ }\href
  {https://doi.org/10.1103/PhysRevD.101.034020} {\bibfield  {journal} {\bibinfo
   {journal} {Phys. Rev. D}\ }\textbf {\bibinfo {volume} {101}},\ \bibinfo
  {pages} {034020} (\bibinfo {year} {2020})},\ \Eprint
  {https://arxiv.org/abs/1807.06566} {arXiv:1807.06566 [hep-lat]} \BibitemShut
  {NoStop}%
\bibitem [{\citenamefont {Alexandrou}\ \emph {et~al.}(2019)\citenamefont
  {Alexandrou}, \citenamefont {Cichy}, \citenamefont {Constantinou},
  \citenamefont {Hadjiyiannakou}, \citenamefont {Jansen}, \citenamefont
  {Scapellato},\ and\ \citenamefont {Steffens}}]{Alexandrou:2019lfo}%
  \BibitemOpen
  \bibfield  {author} {\bibinfo {author} {\bibfnamefont {C.}~\bibnamefont
  {Alexandrou}}, \bibinfo {author} {\bibfnamefont {K.}~\bibnamefont {Cichy}},
  \bibinfo {author} {\bibfnamefont {M.}~\bibnamefont {Constantinou}}, \bibinfo
  {author} {\bibfnamefont {K.}~\bibnamefont {Hadjiyiannakou}}, \bibinfo
  {author} {\bibfnamefont {K.}~\bibnamefont {Jansen}}, \bibinfo {author}
  {\bibfnamefont {A.}~\bibnamefont {Scapellato}},\ and\ \bibinfo {author}
  {\bibfnamefont {F.}~\bibnamefont {Steffens}},\ }\bibfield  {title} {\bibinfo
  {title} {{Systematic uncertainties in parton distribution functions from
  lattice QCD simulations at the physical point}},\ }\href
  {https://doi.org/10.1103/PhysRevD.99.114504} {\bibfield  {journal} {\bibinfo
  {journal} {Phys. Rev. D}\ }\textbf {\bibinfo {volume} {99}},\ \bibinfo
  {pages} {114504} (\bibinfo {year} {2019})},\ \Eprint
  {https://arxiv.org/abs/1902.00587} {arXiv:1902.00587 [hep-lat]} \BibitemShut
  {NoStop}%
\bibitem [{\citenamefont {Ma}\ and\ \citenamefont
  {Qiu}(2018{\natexlab{b}})}]{Ma:2018bpg}%
  \BibitemOpen
  \bibfield  {author} {\bibinfo {author} {\bibfnamefont {Y.-Q.}\ \bibnamefont
  {Ma}}\ and\ \bibinfo {author} {\bibfnamefont {J.-W.}\ \bibnamefont {Qiu}},\
  }\bibfield  {title} {\bibinfo {title} {{Extracting Parton Distribution
  Functions from Lattice QCD Calculations}},\ }\href
  {https://doi.org/10.1103/PhysRevD.98.074021} {\bibfield  {journal} {\bibinfo
  {journal} {Phys. Rev. D}\ }\textbf {\bibinfo {volume} {98}},\ \bibinfo
  {pages} {074021} (\bibinfo {year} {2018}{\natexlab{b}})},\ \Eprint
  {https://arxiv.org/abs/1804.02178} {arXiv:1804.02178 [hep-ph]} \BibitemShut
  {NoStop}%
\bibitem [{\citenamefont {Bouchard}\ \emph {et~al.}(2017)\citenamefont
  {Bouchard}, \citenamefont {Chang}, \citenamefont {Kurth}, \citenamefont
  {Orginos},\ and\ \citenamefont {Walker-Loud}}]{Bouchard:2016heu}%
  \BibitemOpen
  \bibfield  {author} {\bibinfo {author} {\bibfnamefont {C.}~\bibnamefont
  {Bouchard}}, \bibinfo {author} {\bibfnamefont {C.~C.}\ \bibnamefont {Chang}},
  \bibinfo {author} {\bibfnamefont {T.}~\bibnamefont {Kurth}}, \bibinfo
  {author} {\bibfnamefont {K.}~\bibnamefont {Orginos}},\ and\ \bibinfo {author}
  {\bibfnamefont {A.}~\bibnamefont {Walker-Loud}},\ }\bibfield  {title}
  {\bibinfo {title} {{On the Feynman-Hellmann Theorem in Quantum Field Theory
  and the Calculation of Matrix Elements}},\ }\href
  {https://doi.org/10.1103/PhysRevD.96.014504} {\bibfield  {journal} {\bibinfo
  {journal} {Phys. Rev. D}\ }\textbf {\bibinfo {volume} {96}},\ \bibinfo
  {pages} {014504} (\bibinfo {year} {2017})},\ \Eprint
  {https://arxiv.org/abs/1612.06963} {arXiv:1612.06963 [hep-lat]} \BibitemShut
  {NoStop}%
\bibitem [{\citenamefont {Luscher}\ and\ \citenamefont
  {Wolff}(1990)}]{Luscher:1990ux}%
  \BibitemOpen
  \bibfield  {author} {\bibinfo {author} {\bibfnamefont {M.}~\bibnamefont
  {Luscher}}\ and\ \bibinfo {author} {\bibfnamefont {U.}~\bibnamefont
  {Wolff}},\ }\bibfield  {title} {\bibinfo {title} {{How to Calculate the
  Elastic Scattering Matrix in Two-dimensional Quantum Field Theories by
  Numerical Simulation}},\ }\href
  {https://doi.org/10.1016/0550-3213(90)90540-T} {\bibfield  {journal}
  {\bibinfo  {journal} {Nucl. Phys. B}\ }\textbf {\bibinfo {volume} {339}},\
  \bibinfo {pages} {222} (\bibinfo {year} {1990})}\BibitemShut {NoStop}%
\bibitem [{\citenamefont {Blossier}\ \emph {et~al.}(2009)\citenamefont
  {Blossier}, \citenamefont {Della~Morte}, \citenamefont {von Hippel},
  \citenamefont {Mendes},\ and\ \citenamefont {Sommer}}]{Blossier:2009kd}%
  \BibitemOpen
  \bibfield  {author} {\bibinfo {author} {\bibfnamefont {B.}~\bibnamefont
  {Blossier}}, \bibinfo {author} {\bibfnamefont {M.}~\bibnamefont
  {Della~Morte}}, \bibinfo {author} {\bibfnamefont {G.}~\bibnamefont {von
  Hippel}}, \bibinfo {author} {\bibfnamefont {T.}~\bibnamefont {Mendes}},\ and\
  \bibinfo {author} {\bibfnamefont {R.}~\bibnamefont {Sommer}},\ }\bibfield
  {title} {\bibinfo {title} {{On the generalized eigenvalue method for energies
  and matrix elements in lattice field theory}},\ }\href
  {https://doi.org/10.1088/1126-6708/2009/04/094} {\bibfield  {journal}
  {\bibinfo  {journal} {JHEP}\ }\textbf {\bibinfo {volume} {04}}\bibfield
  {number} {\bibinfo  {number} { (2009)},\ \bibinfo {pages} {094}},\ }\Eprint
  {https://arxiv.org/abs/0902.1265} {arXiv:0902.1265 [hep-lat]} \BibitemShut
  {NoStop}%
\bibitem [{\citenamefont {Wagman}(2025)}]{Wagman:2024rid}%
  \BibitemOpen
  \bibfield  {author} {\bibinfo {author} {\bibfnamefont {M.~L.}\ \bibnamefont
  {Wagman}},\ }\bibfield  {title} {\bibinfo {title} {{Lanczos Algorithm, the
  Transfer Matrix, and the Signal-to-Noise Problem}},\ }\href
  {https://doi.org/10.1103/pcvc-734h} {\bibfield  {journal} {\bibinfo
  {journal} {Phys. Rev. Lett.}\ }\textbf {\bibinfo {volume} {134}},\ \bibinfo
  {pages} {241901} (\bibinfo {year} {2025})},\ \Eprint
  {https://arxiv.org/abs/2406.20009} {arXiv:2406.20009 [hep-lat]} \BibitemShut
  {NoStop}%
\bibitem [{\citenamefont {Hackett}\ and\ \citenamefont
  {Wagman}(2025{\natexlab{a}})}]{Hackett:2024xnx}%
  \BibitemOpen
  \bibfield  {author} {\bibinfo {author} {\bibfnamefont {D.~C.}\ \bibnamefont
  {Hackett}}\ and\ \bibinfo {author} {\bibfnamefont {M.~L.}\ \bibnamefont
  {Wagman}},\ }\bibfield  {title} {\bibinfo {title} {{Lanczos algorithm for
  lattice QCD matrix elements}},\ }\href {https://doi.org/10.1103/zjzt-rv86}
  {\bibfield  {journal} {\bibinfo  {journal} {Phys. Rev. D}\ }\textbf {\bibinfo
  {volume} {112}},\ \bibinfo {pages} {054506} (\bibinfo {year}
  {2025}{\natexlab{a}})},\ \Eprint {https://arxiv.org/abs/2407.21777}
  {arXiv:2407.21777 [hep-lat]} \BibitemShut {NoStop}%
\bibitem [{\citenamefont {Hackett}\ and\ \citenamefont
  {Wagman}(2025{\natexlab{b}})}]{Hackett:2024nbe}%
  \BibitemOpen
  \bibfield  {author} {\bibinfo {author} {\bibfnamefont {D.~C.}\ \bibnamefont
  {Hackett}}\ and\ \bibinfo {author} {\bibfnamefont {M.~L.}\ \bibnamefont
  {Wagman}},\ }\bibfield  {title} {\bibinfo {title} {{Block Lanczos algorithm
  for lattice QCD spectroscopy and matrix elements}},\ }\href
  {https://doi.org/10.1103/fp74-q35q} {\bibfield  {journal} {\bibinfo
  {journal} {Phys. Rev. D}\ }\textbf {\bibinfo {volume} {112}},\ \bibinfo
  {pages} {014514} (\bibinfo {year} {2025}{\natexlab{b}})},\ \Eprint
  {https://arxiv.org/abs/2412.04444} {arXiv:2412.04444 [hep-lat]} \BibitemShut
  {NoStop}%
\bibitem [{\citenamefont {Chen}\ \emph {et~al.}(2026)\citenamefont {Chen} \emph
  {et~al.}}]{Chen:2025cxr}%
  \BibitemOpen
  \bibfield  {author} {\bibinfo {author} {\bibfnamefont {J.-W.}\ \bibnamefont
  {Chen}} \emph {et~al.},\ }\bibfield  {title} {\bibinfo {title}
  {{Large-momentum effective theory{\textquoteright}s asymptotic extrapolation
  vs the inverse problem}},\ }\href {https://doi.org/10.1103/fflw-qpcc}
  {\bibfield  {journal} {\bibinfo  {journal} {Phys. Rev. D}\ }\textbf {\bibinfo
  {volume} {113}},\ \bibinfo {pages} {014509} (\bibinfo {year} {2026})},\
  \Eprint {https://arxiv.org/abs/2505.14619} {arXiv:2505.14619 [hep-lat]}
  \BibitemShut {NoStop}%
\bibitem [{\citenamefont {Jay}\ and\ \citenamefont {Neil}(2021)}]{Jay:2020ghd}%
  \BibitemOpen
  \bibfield  {author} {\bibinfo {author} {\bibfnamefont {W.~I.}\ \bibnamefont
  {Jay}}\ and\ \bibinfo {author} {\bibfnamefont {E.~T.}\ \bibnamefont {Neil}},\
  }\bibfield  {title} {\bibinfo {title} {{Bayesian model averaging for analysis
  of lattice field theory results}},\ }\href
  {https://doi.org/10.1103/PhysRevD.103.114502} {\bibfield  {journal} {\bibinfo
   {journal} {Phys. Rev. D}\ }\textbf {\bibinfo {volume} {103}},\ \bibinfo
  {pages} {114502} (\bibinfo {year} {2021})},\ \Eprint
  {https://arxiv.org/abs/2008.01069} {arXiv:2008.01069 [hep-lat]} \BibitemShut
  {NoStop}%
\bibitem [{\citenamefont {Neil}\ and\ \citenamefont
  {Sitison}(2024)}]{Neil:2022jmn}%
  \BibitemOpen
  \bibfield  {author} {\bibinfo {author} {\bibfnamefont {E.~T.}\ \bibnamefont
  {Neil}}\ and\ \bibinfo {author} {\bibfnamefont {J.~W.}\ \bibnamefont
  {Sitison}},\ }\bibfield  {title} {\bibinfo {title} {{Improved information
  criteria for Bayesian model averaging in lattice field theory}},\ }\href
  {https://doi.org/10.1103/PhysRevD.109.014510} {\bibfield  {journal} {\bibinfo
   {journal} {Phys. Rev. D}\ }\textbf {\bibinfo {volume} {109}},\ \bibinfo
  {pages} {014510} (\bibinfo {year} {2024})},\ \Eprint
  {https://arxiv.org/abs/2208.14983} {arXiv:2208.14983 [hep-lat]} \BibitemShut
  {NoStop}%
\bibitem [{\citenamefont {Zhang}(2023)}]{Zhang:2023wea}%
  \BibitemOpen
  \bibfield  {author} {\bibinfo {author} {\bibfnamefont {J.-H.}\ \bibnamefont
  {Zhang}},\ }\href@noop {} {\bibinfo {title} {{Double Parton Distributions
  from Euclidean Lattice}}} (\bibinfo {year} {2023}),\ \Eprint
  {https://arxiv.org/abs/2304.12481} {arXiv:2304.12481 [hep-ph]} \BibitemShut
  {NoStop}%
\bibitem [{\citenamefont {Jaarsma}\ \emph {et~al.}(2023)\citenamefont
  {Jaarsma}, \citenamefont {Rahn},\ and\ \citenamefont
  {Waalewijn}}]{Jaarsma:2023woo}%
  \BibitemOpen
  \bibfield  {author} {\bibinfo {author} {\bibfnamefont {M.}~\bibnamefont
  {Jaarsma}}, \bibinfo {author} {\bibfnamefont {R.}~\bibnamefont {Rahn}},\ and\
  \bibinfo {author} {\bibfnamefont {W.~J.}\ \bibnamefont {Waalewijn}},\
  }\bibfield  {title} {\bibinfo {title} {{Towards double parton distributions
  from first principles using Large Momentum Effective Theory}},\ }\href
  {https://doi.org/10.1007/JHEP12(2023)014} {\bibfield  {journal} {\bibinfo
  {journal} {JHEP}\ }\textbf {\bibinfo {volume} {12}}\bibfield  {number}
  {\bibinfo  {number} { (2023)},\ \bibinfo {pages} {014}},\ }\Eprint
  {https://arxiv.org/abs/2305.09716} {arXiv:2305.09716 [hep-ph]} \BibitemShut
  {NoStop}%
\bibitem [{\citenamefont {Reitinger}\ \emph {et~al.}(2024)\citenamefont
  {Reitinger}, \citenamefont {Zimmermann}, \citenamefont {Diehl},\ and\
  \citenamefont {Sch{\"a}fer}}]{Reitinger:2024ulw}%
  \BibitemOpen
  \bibfield  {author} {\bibinfo {author} {\bibfnamefont {D.}~\bibnamefont
  {Reitinger}}, \bibinfo {author} {\bibfnamefont {C.}~\bibnamefont
  {Zimmermann}}, \bibinfo {author} {\bibfnamefont {M.}~\bibnamefont {Diehl}},\
  and\ \bibinfo {author} {\bibfnamefont {A.}~\bibnamefont {Sch{\"a}fer}},\
  }\bibfield  {title} {\bibinfo {title} {{Double parton distributions with
  flavor interference from lattice QCD}},\ }\href
  {https://doi.org/10.1007/JHEP04(2024)087} {\bibfield  {journal} {\bibinfo
  {journal} {JHEP}\ }\textbf {\bibinfo {volume} {04}}\bibfield  {number}
  {\bibinfo  {number} { (2024)},\ \bibinfo {pages} {087}},\ }\Eprint
  {https://arxiv.org/abs/2401.14855} {arXiv:2401.14855 [hep-lat]} \BibitemShut
  {NoStop}%
\bibitem [{\citenamefont {Diehl}\ \emph {et~al.}(2026)\citenamefont {Diehl},
  \citenamefont {Grocholski}, \citenamefont {Reitinger}, \citenamefont
  {Sch{\"a}fer},\ and\ \citenamefont {Zimmermann}}]{Diehl:2025kcr}%
  \BibitemOpen
  \bibfield  {author} {\bibinfo {author} {\bibfnamefont {M.}~\bibnamefont
  {Diehl}}, \bibinfo {author} {\bibfnamefont {O.}~\bibnamefont {Grocholski}},
  \bibinfo {author} {\bibfnamefont {D.}~\bibnamefont {Reitinger}}, \bibinfo
  {author} {\bibfnamefont {A.}~\bibnamefont {Sch{\"a}fer}},\ and\ \bibinfo
  {author} {\bibfnamefont {C.}~\bibnamefont {Zimmermann}},\ }\bibfield  {title}
  {\bibinfo {title} {{Extracting Mellin moments of double parton distributions
  from lattice data}},\ }\href {https://doi.org/10.1007/JHEP05(2026)004}
  {\bibfield  {journal} {\bibinfo  {journal} {JHEP}\ }\textbf {\bibinfo
  {volume} {05}}\bibfield  {number} {\bibinfo  {number} { (2026)},\ \bibinfo
  {pages} {004}},\ }\Eprint {https://arxiv.org/abs/2511.15546}
  {arXiv:2511.15546 [hep-lat]} \BibitemShut {NoStop}%
\bibitem [{\citenamefont {Yao}\ \emph {et~al.}(2023)\citenamefont {Yao} \emph
  {et~al.}}]{LatticeParton:2022xsd}%
  \BibitemOpen
  \bibfield  {author} {\bibinfo {author} {\bibfnamefont {F.}~\bibnamefont
  {Yao}} \emph {et~al.} (\bibinfo {collaboration} {Lattice Parton}),\
  }\bibfield  {title} {\bibinfo {title} {{Nucleon Transversity Distribution in
  the Continuum and Physical Mass Limit from Lattice QCD}},\ }\href
  {https://doi.org/10.1103/PhysRevLett.131.261901} {\bibfield  {journal}
  {\bibinfo  {journal} {Phys. Rev. Lett.}\ }\textbf {\bibinfo {volume} {131}},\
  \bibinfo {pages} {261901} (\bibinfo {year} {2023})},\ \Eprint
  {https://arxiv.org/abs/2208.08008} {arXiv:2208.08008 [hep-lat]} \BibitemShut
  {NoStop}%
\bibitem [{\citenamefont {Chen}\ \emph {et~al.}(2025)\citenamefont {Chen},
  \citenamefont {Zeng}, \citenamefont {Dong}, \citenamefont {Liu},
  \citenamefont {Shen}, \citenamefont {Sun}, \citenamefont {Xiong},
  \citenamefont {Yang}, \citenamefont {Yao}, \citenamefont {Zhang},\ and\
  \citenamefont {Zhong}}]{Chen:2025gluon}%
  \BibitemOpen
  \bibfield  {author} {\bibinfo {author} {\bibfnamefont {C.}~\bibnamefont
  {Chen}}, \bibinfo {author} {\bibfnamefont {C.}~\bibnamefont {Zeng}}, \bibinfo
  {author} {\bibfnamefont {H.}~\bibnamefont {Dong}}, \bibinfo {author}
  {\bibfnamefont {L.}~\bibnamefont {Liu}}, \bibinfo {author} {\bibfnamefont
  {X.}~\bibnamefont {Shen}}, \bibinfo {author} {\bibfnamefont {P.}~\bibnamefont
  {Sun}}, \bibinfo {author} {\bibfnamefont {X.}~\bibnamefont {Xiong}}, \bibinfo
  {author} {\bibfnamefont {Y.-B.}\ \bibnamefont {Yang}}, \bibinfo {author}
  {\bibfnamefont {F.}~\bibnamefont {Yao}}, \bibinfo {author} {\bibfnamefont
  {J.-H.}\ \bibnamefont {Zhang}},\ and\ \bibinfo {author} {\bibfnamefont
  {S.}~\bibnamefont {Zhong}},\ }\href@noop {} {\bibinfo {title} {{Unpolarized
  gluon PDF of the nucleon from lattice QCD at physical point in the continuum
  limit}}} (\bibinfo {year} {2025}),\ \Eprint
  {https://arxiv.org/abs/2510.26425} {arXiv:2510.26425 [hep-lat]} \BibitemShut
  {NoStop}%
\bibitem [{\citenamefont {Hu}\ \emph {et~al.}(2024)\citenamefont {Hu},
  \citenamefont {Hu}, \citenamefont {Wang}, \citenamefont {Gong}, \citenamefont
  {Liu}, \citenamefont {Sun}, \citenamefont {Sun}, \citenamefont {Wang},
  \citenamefont {Yang},\ and\ \citenamefont {Zhao}}]{Hu:2023qmc}%
  \BibitemOpen
  \bibfield  {author} {\bibinfo {author} {\bibfnamefont {Z.-C.}\ \bibnamefont
  {Hu}}, \bibinfo {author} {\bibfnamefont {B.-L.}\ \bibnamefont {Hu}}, \bibinfo
  {author} {\bibfnamefont {J.-H.}\ \bibnamefont {Wang}}, \bibinfo {author}
  {\bibfnamefont {M.}~\bibnamefont {Gong}}, \bibinfo {author} {\bibfnamefont
  {L.}~\bibnamefont {Liu}}, \bibinfo {author} {\bibfnamefont {P.}~\bibnamefont
  {Sun}}, \bibinfo {author} {\bibfnamefont {W.}~\bibnamefont {Sun}}, \bibinfo
  {author} {\bibfnamefont {W.}~\bibnamefont {Wang}}, \bibinfo {author}
  {\bibfnamefont {Y.-B.}\ \bibnamefont {Yang}},\ and\ \bibinfo {author}
  {\bibfnamefont {D.-J.}\ \bibnamefont {Zhao}},\ }\bibfield  {title} {\bibinfo
  {title} {{Quark masses and low energy constants in the continuum from the
  tadpole improved clover ensembles}},\ }\href
  {https://doi.org/10.1103/PhysRevD.109.054507} {\bibfield  {journal} {\bibinfo
   {journal} {Phys. Rev. D}\ }\textbf {\bibinfo {volume} {109}},\ \bibinfo
  {pages} {054507} (\bibinfo {year} {2024})},\ \Eprint
  {https://arxiv.org/abs/2310.00814} {arXiv:2310.00814 [hep-lat]} \BibitemShut
  {NoStop}%
\bibitem [{\citenamefont {Du}\ \emph {et~al.}(2025)\citenamefont {Du},
  \citenamefont {Hu}, \citenamefont {Cai}, \citenamefont {Chen}, \citenamefont
  {Ding}, \citenamefont {Jiang}, \citenamefont {Liu}, \citenamefont {Liu},
  \citenamefont {Meng}, \citenamefont {Sun}, \citenamefont {Wang},
  \citenamefont {Yang},\ and\ \citenamefont {Zhao}}]{CLQCD:2024tad}%
  \BibitemOpen
  \bibfield  {author} {\bibinfo {author} {\bibfnamefont {H.-Y.}\ \bibnamefont
  {Du}}, \bibinfo {author} {\bibfnamefont {B.}~\bibnamefont {Hu}}, \bibinfo
  {author} {\bibfnamefont {M.}~\bibnamefont {Cai}}, \bibinfo {author}
  {\bibfnamefont {Y.}~\bibnamefont {Chen}}, \bibinfo {author} {\bibfnamefont
  {H.-T.}\ \bibnamefont {Ding}}, \bibinfo {author} {\bibfnamefont
  {X.}~\bibnamefont {Jiang}}, \bibinfo {author} {\bibfnamefont
  {C.}~\bibnamefont {Liu}}, \bibinfo {author} {\bibfnamefont {L.}~\bibnamefont
  {Liu}}, \bibinfo {author} {\bibfnamefont {Y.}~\bibnamefont {Meng}}, \bibinfo
  {author} {\bibfnamefont {P.}~\bibnamefont {Sun}}, \bibinfo {author}
  {\bibfnamefont {J.-H.}\ \bibnamefont {Wang}}, \bibinfo {author}
  {\bibfnamefont {Y.-B.}\ \bibnamefont {Yang}},\ and\ \bibinfo {author}
  {\bibfnamefont {D.-J.}\ \bibnamefont {Zhao}} (\bibinfo {collaboration}
  {CLQCD}),\ }\bibfield  {title} {\bibinfo {title} {{Charmed meson masses and
  decay constants in the continuum from the tadpole improved clover
  ensembles}},\ }\href {https://doi.org/10.1103/PhysRevD.111.054504} {\bibfield
   {journal} {\bibinfo  {journal} {Phys. Rev. D}\ }\textbf {\bibinfo {volume}
  {111}},\ \bibinfo {pages} {054504} (\bibinfo {year} {2025})},\ \Eprint
  {https://arxiv.org/abs/2408.03548} {arXiv:2408.03548 [hep-lat]} \BibitemShut
  {NoStop}%
\bibitem [{\citenamefont {Ma}\ \emph {et~al.}(2025)\citenamefont {Ma} \emph
  {et~al.}}]{LPC:2025spt}%
  \BibitemOpen
  \bibfield  {author} {\bibinfo {author} {\bibfnamefont {L.}~\bibnamefont {Ma}}
  \emph {et~al.} (\bibinfo {collaboration} {Lattice Parton}),\ }\bibfield
  {title} {\bibinfo {title} {{Quark transverse spin-momentum correlation of the
  nucleon from lattice QCD: the Boer-Mulders function}},\ }\href
  {https://doi.org/10.1007/JHEP08(2025)086} {\bibfield  {journal} {\bibinfo
  {journal} {JHEP}\ }\textbf {\bibinfo {volume} {08}}\bibfield  {number}
  {\bibinfo  {number} { (2025)},\ \bibinfo {pages} {086}},\ }\Eprint
  {https://arxiv.org/abs/2502.11807} {arXiv:2502.11807 [hep-lat]} \BibitemShut
  {NoStop}%
\bibitem [{\citenamefont {Zhang}\ \emph {et~al.}(2019)\citenamefont {Zhang},
  \citenamefont {Ji}, \citenamefont {Sch{\"a}fer}, \citenamefont {Wang},\ and\
  \citenamefont {Zhao}}]{Zhang:2018diq}%
  \BibitemOpen
  \bibfield  {author} {\bibinfo {author} {\bibfnamefont {J.-H.}\ \bibnamefont
  {Zhang}}, \bibinfo {author} {\bibfnamefont {X.}~\bibnamefont {Ji}}, \bibinfo
  {author} {\bibfnamefont {A.}~\bibnamefont {Sch{\"a}fer}}, \bibinfo {author}
  {\bibfnamefont {W.}~\bibnamefont {Wang}},\ and\ \bibinfo {author}
  {\bibfnamefont {S.}~\bibnamefont {Zhao}},\ }\bibfield  {title} {\bibinfo
  {title} {{Accessing Gluon Parton Distributions in Large Momentum Effective
  Theory}},\ }\href {https://doi.org/10.1103/PhysRevLett.122.142001} {\bibfield
   {journal} {\bibinfo  {journal} {Phys. Rev. Lett.}\ }\textbf {\bibinfo
  {volume} {122}},\ \bibinfo {pages} {142001} (\bibinfo {year} {2019})},\
  \Eprint {https://arxiv.org/abs/1808.10824} {arXiv:1808.10824 [hep-ph]}
  \BibitemShut {NoStop}%
\bibitem [{\citenamefont {Chen}\ \emph {et~al.}(2020)\citenamefont {Chen},
  \citenamefont {Lin},\ and\ \citenamefont {Zhang}}]{Chen:2019lcm}%
  \BibitemOpen
  \bibfield  {author} {\bibinfo {author} {\bibfnamefont {J.-W.}\ \bibnamefont
  {Chen}}, \bibinfo {author} {\bibfnamefont {H.-W.}\ \bibnamefont {Lin}},\ and\
  \bibinfo {author} {\bibfnamefont {J.-H.}\ \bibnamefont {Zhang}},\ }\bibfield
  {title} {\bibinfo {title} {{Pion generalized parton distribution from lattice
  QCD}},\ }\href {https://doi.org/10.1016/j.nuclphysb.2020.114940} {\bibfield
  {journal} {\bibinfo  {journal} {Nucl. Phys. B}\ }\textbf {\bibinfo {volume}
  {952}},\ \bibinfo {pages} {114940} (\bibinfo {year} {2020})},\ \Eprint
  {https://arxiv.org/abs/1904.12376} {arXiv:1904.12376 [hep-lat]} \BibitemShut
  {NoStop}%
\bibitem [{\citenamefont {Lin}(2023)}]{Lin:2023gxz}%
  \BibitemOpen
  \bibfield  {author} {\bibinfo {author} {\bibfnamefont {H.-W.}\ \bibnamefont
  {Lin}},\ }\bibfield  {title} {\bibinfo {title} {{Pion valence-quark
  generalized parton distribution at physical pion mass}},\ }\href
  {https://doi.org/10.1016/j.physletb.2023.138181} {\bibfield  {journal}
  {\bibinfo  {journal} {Phys. Lett. B}\ }\textbf {\bibinfo {volume} {846}},\
  \bibinfo {pages} {138181} (\bibinfo {year} {2023})},\ \Eprint
  {https://arxiv.org/abs/2310.10579} {arXiv:2310.10579 [hep-lat]} \BibitemShut
  {NoStop}%
\bibitem [{\citenamefont {Ding}\ \emph {et~al.}(2025)\citenamefont {Ding},
  \citenamefont {Gao}, \citenamefont {Mukherjee}, \citenamefont {Petreczky},
  \citenamefont {Shi}, \citenamefont {Syritsyn},\ and\ \citenamefont
  {Zhao}}]{Ding:2024saz}%
  \BibitemOpen
  \bibfield  {author} {\bibinfo {author} {\bibfnamefont {H.-T.}\ \bibnamefont
  {Ding}}, \bibinfo {author} {\bibfnamefont {X.}~\bibnamefont {Gao}}, \bibinfo
  {author} {\bibfnamefont {S.}~\bibnamefont {Mukherjee}}, \bibinfo {author}
  {\bibfnamefont {P.}~\bibnamefont {Petreczky}}, \bibinfo {author}
  {\bibfnamefont {Q.}~\bibnamefont {Shi}}, \bibinfo {author} {\bibfnamefont
  {S.}~\bibnamefont {Syritsyn}},\ and\ \bibinfo {author} {\bibfnamefont
  {Y.}~\bibnamefont {Zhao}},\ }\bibfield  {title} {\bibinfo {title}
  {{Three-dimensional imaging of pion using lattice QCD: generalized parton
  distributions}},\ }\href {https://doi.org/10.1007/JHEP02(2025)056} {\bibfield
   {journal} {\bibinfo  {journal} {JHEP}\ }\textbf {\bibinfo {volume}
  {02}}\bibfield  {number} {\bibinfo  {number} { (2025)},\ \bibinfo {pages}
  {056}},\ }\Eprint {https://arxiv.org/abs/2407.03516} {arXiv:2407.03516
  [hep-lat]} \BibitemShut {NoStop}%
\bibitem [{\citenamefont {Zhao}\ \emph {et~al.}(2026)\citenamefont {Zhao} \emph
  {et~al.}}]{Zhao:toappear}%
  \BibitemOpen
  \bibfield  {author} {\bibinfo {author} {\bibfnamefont {D.-J.}\ \bibnamefont
  {Zhao}} \emph {et~al.},\ }\href@noop {} {\bibinfo {title} {{In preparation}}}
  (\bibinfo {year} {2026})\BibitemShut {NoStop}%
\bibitem [{\citenamefont {Boer}\ and\ \citenamefont
  {Mulders}(1998)}]{Boer:1997nt}%
  \BibitemOpen
  \bibfield  {author} {\bibinfo {author} {\bibfnamefont {D.}~\bibnamefont
  {Boer}}\ and\ \bibinfo {author} {\bibfnamefont {P.~J.}\ \bibnamefont
  {Mulders}},\ }\bibfield  {title} {\bibinfo {title} {{Time-reversal odd
  distribution functions in leptoproduction}},\ }\href
  {https://doi.org/10.1103/PhysRevD.57.5780} {\bibfield  {journal} {\bibinfo
  {journal} {Phys. Rev. D}\ }\textbf {\bibinfo {volume} {57}},\ \bibinfo
  {pages} {5780} (\bibinfo {year} {1998})},\ \Eprint
  {https://arxiv.org/abs/hep-ph/9711485} {arXiv:hep-ph/9711485 [hep-ph]}
  \BibitemShut {NoStop}%
\bibitem [{\citenamefont {Mulders}\ and\ \citenamefont
  {Tangerman}(1996)}]{Mulders:1995dh}%
  \BibitemOpen
  \bibfield  {author} {\bibinfo {author} {\bibfnamefont {P.~J.}\ \bibnamefont
  {Mulders}}\ and\ \bibinfo {author} {\bibfnamefont {R.~D.}\ \bibnamefont
  {Tangerman}},\ }\bibfield  {title} {\bibinfo {title} {{The complete tree
  level result up to order 1/Q for polarized deep inelastic leptoproduction}},\
  }\href {https://doi.org/10.1016/0550-3213(95)00632-X} {\bibfield  {journal}
  {\bibinfo  {journal} {Nucl. Phys. B}\ }\textbf {\bibinfo {volume} {461}},\
  \bibinfo {pages} {197} (\bibinfo {year} {1996})},\ \bibinfo {note} {[Erratum:
  Nucl. Phys. B 484 (1997) 538]},\ \Eprint
  {https://arxiv.org/abs/hep-ph/9510301} {arXiv:hep-ph/9510301 [hep-ph]}
  \BibitemShut {NoStop}%
\end{thebibliography}%

\end{document}